\documentclass[12pt]{article}
\usepackage[utf8]{inputenc}
\usepackage{graphicx,psfrag,epsf}
\usepackage{enumerate}
\usepackage{natbib}
\usepackage[margin = 1in]{geometry}
\usepackage{color}
\usepackage{amsmath,amssymb}
\usepackage{bm}
\usepackage{listings}
\usepackage[svgnames]{xcolor}
\usepackage{natbib}
\usepackage{setspace}
\usepackage{url}

\renewenvironment{abstract}
 {\small
  \begin{center}
  \bfseries \abstractname\vspace{-0mm}\vspace{0pt}
  \end{center}
  \list{}{
    \setlength{\leftmargin}{1cm}%
    \setlength{\rightmargin}{\leftmargin}%
  }%
  \item\relax}
 {\endlist}

\def\spacingset#1{\renewcommand{\baselinestretch}%
{#1}\small\normalsize} \spacingset{1}

\renewcommand{\baselinestretch}{1.5}

\begin{document}

\title{\bf \LARGE Smaller $p$-values in genomics studies using distilled historical information}

\author{Jordan G. Bryan, Peter D. Hoff\\[4pt]
\textit{Department of Statistical Science, Duke University, Durham, NC 27708}\\[1pt]
{jordan.bryan@duke.edu, peter.hoff@duke.edu}}

\markboth%
{J. G. Bryan and P. D. Hoff}
{Smaller $p$-values in genomics studies}

\maketitle

\begin{abstract}
{Medical research institutions have generated massive amounts of biological data by genetically profiling hundreds of cancer cell lines. In parallel, academic biology labs have conducted genetic screens on small numbers of cancer cell lines under custom experimental conditions. In order to share information between these two approaches to scientific discovery, this article proposes a ``frequentist assisted by Bayes" (FAB) procedure for hypothesis testing that allows historical information from massive genomics datasets to increase the power of hypothesis tests in specialized studies. The exchange of information takes place through a novel probability model for multimodal genomics data, which distills historical information pertaining to cancer cell lines and genes across a wide variety of experimental contexts. If the relevance of the historical information for a given study is high, then the resulting FAB tests can be more powerful than the corresponding classical tests. If the relevance is low, then the FAB tests yield as many discoveries as the classical tests. Simulations and practical investigations demonstrate that the FAB testing procedure can increase the number of effects discovered in genomics studies while still maintaining strict control of type I error and false discovery rates.}  
\end{abstract}

\section{Introduction}
\label{sec1}

In recent years, non-profit research organizations have initiated large-scale   
efforts to create genetic maps of cancer \citep{tsherniak_defining_2017, meyers_computational_2017, mcfarland_improved_2018, iorio_landscape_2016, behan_prioritization_2019, ghandi_next-generation_2019}. The data produced by these projects measure different modalities of the cancer genome. For instance, RNAseq data measure how much RNA is being expressed from a particular gene in a given cancer cell line. CRISPR dependency data measure the extent to which the proliferation of a cancer cell line is affected upon deletion of a given gene. DNA sequencing data record the presence or absence of aberrant nucleotide sequences in the coding regions of genes in cancer cell lines.

The purpose of exploratory \textit{data-driven} screens like these is to discover patterns in the relationships among cancer cell lines and among genes, patterns which will hopefully yield new biological insights. However, many biological questions cannot be answered without \textit{hypothesis-driven} experiments, which complement the scattershot approach of data-driven screens by testing the effects of targeted interventions on the baseline state of cellular models of cancer. For example, the CRISPR-Cas9 gene editing technology has been used in conjunction with traditional treatment-control experimental designs to successfully uncover and validate the functional role of several genes \citep{kory_sfxn1_2018, birsoy_essential_2015, pusapati_crispr_2018, liao_genetic_2017}. 

An open question in modern biology is how to properly inform  
hypothesis-driven experimenters with the information available in massive genomics datasets created by data-driven screens.
The most conspicuous efforts to deliver genomics information to biologists come in the form of online data portals. Online portals allow biologists to browse and visualize a corpus of publicly available genomics data without the need for computing expertise. For this reason, portals are invaluable tools for strengthening biological arguments, sleuthing in an ad-hoc manner to find connections to genetic mechanisms of interest, and generating new hypotheses. However, data portals do not provide a method for directly 
incorporating information from the corpus into the statistical evaluation of new hypotheses.  
Modern statistical methods applicable to analyzing large genomics datasets are also potentially ill-suited to the needs of hypothesis-driven experiments. While the technologies and methodologies employed by a modern working biologist are complex, the statistical tests necessary to extract insights from their data need not be. 

The current state of affairs in biology therefore demands 
a framework for 
(1) distilling information from a corpus of multimodal genomics data, 
(2) quantifying the relevance of this information to specific hypothesis-driven experiments, and
(3) incorporating relevant information from the corpus into statistical evaluations of specific experimental hypotheses, while maintaining error rate controls. 
This article proposes such a framework that combines the flexibility of 
Bayesian statistical modeling with the statistical guarantees of 
frequentist hypothesis testing procedures. 
Specifically, the proposal is to first distill information 
from a historical corpus with a 
probabilistic tensor factorization model, which quantifies the relationships among a set of biological entities measured by different genomics modalities. Inference from this tensor probability model creates a numerical feature profile for each entity, which encodes historical information distilled from all modalities.
Next, an empirical Bayes procedure is used to  
assess the relevance of these features to data from a new hypothesis-driven study. 
Finally, the relevant information is used to construct hypothesis tests and $p$-values for biological effects measured by the new experiment. This is accomplished by using the recently-developed ``frequentist, assisted by Bayes'' (FAB) hypothesis testing framework described by \cite{hoff_smaller_2019}, which is based on frequentist testing procedures that are optimal with respect to historical information. As a result, if the historical information is highly relevant to the new experimental data, the FAB tests will have higher power than the corresponding classical tests that do not make use of historical information. Conversely, the proposed empirical Bayes procedure generates 
FAB tests whose 
power 
approaches that of classical tests when the historical information is irrelevant. 
The FAB tests retain type I error rate and false discovery rate (FDR)
guarantees whether or not 
the historical information is relevant to the new modality in the study at hand. Hence, the FAB testing procedure described here may be applied with confidence to studies performed with any number of measurement types or experimental designs, even when the utility of the historical information is unknown.

The remainder of the article is organized as follows: Section \ref{fabmod} describes the FAB testing procedure in the context of hypothesis-driven genomics experiments where historical information is available for multiple biological effects. Section \ref{hist_rel} presents an empirical Bayes method for quantifying the relevance of the historical information to a new set of experimental data. Section \ref{dpi} details the probabilistic model for distilling information from multimodal genomics datasets, which makes the FAB procedure applicable to practical settings where historical information for the biological effects under study is not otherwise available. Finally, Sections \ref{simstud} and \ref{examples} describe results from the application of the FAB tests to simulated and real datasets. The simulation studies in Section \ref{simstud} empirically validate the FAB procedure's theoretical type I error properties under null scenarios and showcase its type II behavior under non-null scenarios. Section \ref{examples} presents a comprehensive catalogue of practical settings in which the FAB procedure can be used to discover more biological effects than the classical procedure. The examples presented include CRISPR screens, cell viability profiling for drug repurposing, differential expression analysis, differential dependency analysis, and modifier screens targeted towards specific genetic interactions. 

\section{Methods}
\label{sec2}

Advances in modern sequencing technologies have 
dramatically increased the amount of data produced in
hypothesis-driven biological experiments. 
Analysis of data from such 
experiments
may involve 
hundreds or thousands of statistical hypothesis tests, each corresponding to a biological effect of interest. The following sections describe a procedure by which the presence of multiple hypotheses, often a statistical challenge in the analysis of genomics data, creates an opportunity for increasing the power of hypothesis tests by using historical information. The testing procedure described maintains frequentist type I error guarantees and adaptively estimates the relevance of the historical information to a new set of experimental data so that type II error approaches that of classical tests when the relevance of the historical information is low. When historical information for each of thousands of biological effects under investigation in a particular study is not readily available, a generalization of the probability model underlying the testing procedure may be used to distill that information from multimodal genomics data. The result is a two-stage estimation procedure, which has the potential to increase the number of discoveries made in a wide variety of genomics studies.

\subsection{FAB hypothesis tests for genomics data}\label{fabmod}

Hypothesis-driven genomics experiments are often designed to evaluate a null hypothesis $H_j : \theta_j = 0$ for each of many biological effects $\theta_j,~ j = 1, \dots, M$. In many cases, the data used to evaluate these hypotheses 
consist of independent, approximately normally distributed estimates $\hat\theta_1,\ldots, \hat\theta_M$ of 
the effects $\theta_1,\ldots, \theta_M$,
as well as an estimate 
of the variance of each $\hat\theta_j$. Specifically, 
it is often assumed that 
\begin{align} 
\hat\theta_j & \sim N ( \theta_j , \sigma_j^2/c_j)  
\end{align}
independently for $j=1,\ldots, M$, where each $c_j$ is a known constant. 
Given an estimator $\hat{\sigma}_j^2$ of $\sigma_j^2$, 
the statistic 
$T_j = \sqrt{c_j} \hat \theta_j/\hat\sigma_j$
 has approximately a standard normal 
or $t$-distribution under the null hypothesis $H_j:\theta_j=0$. 
In what follows, $t$-distributions are assumed 
as these are exact in several cases. For example, 
if the data for effect $j$ consist of a sample 
$Y_{1j},\ldots, Y_{nj} \sim $ i.i.d.\ $N(\theta_j,\sigma_j^2)$, then setting
$\hat\theta_j$ to be the sample mean, $\hat\sigma_j^2$ to be the the pooled sample estimate of $\sigma_j^2$, and 
$c_j=n_j$, 
each $T_j$ is exactly $t$-distributed under the hypothesis $H_j$. 

Under this sampling model, the statistical significance of the effect $\theta_{j}$ is typically evaluated using a classical two-tailed $t$-test, which tests $H_j : \theta_{j} = 0$ versus $K_j : \theta_{j} \neq 0$. With some algebraic manipulations, it can be shown that the $p$-value for the two-tailed $t$-test has the following form:
\begin{equation}
    p_j = 1 - \left| F_\nu(T_j) - F_\nu(-T_j)\right|
\end{equation}
where $T_j =  
\sqrt{c_j} \hat\theta_j/\hat\sigma_j$, $F_\nu$ is the cumulative distribution function of the $t$-distribution with $\nu$ degrees of freedom, and $\hat{\sigma_j}^2$ is an estimator of $\sigma_j^2$ for which $\nu \hat{\sigma_j}^2 / \sigma_j^2 \sim \chi_\nu^2$. Under the null hypothesis $H_j$, $p_j$ is uniformly distributed in the interval $(0, 1)$, so the type I error rate of the test may be controlled at level $\alpha$ by rejecting the null when $p_j < \alpha$.

A useful mathematical fact first noted by \cite{hoff_smaller_2019} is that for any value of $b_j$, the following quantity is also uniformly distributed under $H_j$:
\begin{equation}
    p^{\text{FAB}}_{j} = 1 - \left| F_{\nu}\left(T_j + b_{j} \right) - F_{\nu} \left( -T_j \right) \right|.
    \label{eq:p_fab}
\end{equation}
Hypothesis tests based on $p^{\text{FAB}}_{j}$ therefore have the same type I error rates as those based on the classical $p$-value. However, 
under an alternative value of $\theta_j$ with the same sign as $b_j$, 
$p^{\text{FAB}}_{j}$ will be stochastically smaller than $p_j$. Therefore, careful selection of $b_j$ can 
lead to a more powerful hypothesis test than the classical two-sided $t$-test. If $\theta_j$ is positive, then the FAB $p$-value approaches the $p$-value from a one-sided test against the alternative $\theta_j > 0$ as $b_j \rightarrow \infty$. Likewise, if $\theta_j$ is negative, the FAB $p$-value approches that of a one-sided test against the alternative $\theta_j < 0$ as $b_j \rightarrow -\infty$. The FAB $p$-value can therefore be as small as half the classical $p$-value. However, the FAB $p$-value can be larger than the classical $p$-value for values of $\theta_j$ and $b_j$ with opposite sign.

A principled approach to using historical data to inform the choice of 
$b_j$ is to encode information about $\theta_j$ with a 
prior distribution, say $\theta_j \sim N(m_j, v_j)$. Here, the quantities $m_j$ and $v_j$ are historically-informed location and scale parameters for the biological effect $\theta_j$.
As shown in \cite{hoff_smaller_2019}, if $\sigma_j^2$ were known,
the value of $b_j$ that 
maximizes the prior expected power of any test based on $p_j^{\text{FAB}}$ is 
\begin{equation}
    b_j^\text{OPT} = 2 m_j \sigma_j / \sqrt{c_j} v_j. 
\end{equation} 
In practice, $\sigma_j^2$ is not known and so must be replaced with an 
estimator, in which case the corresponding value of $b_j$ is only 
approximately (asymptotically) optimal, in terms of prior 
expected power.

\subsection{Assessing relevance of historical information}\label{hist_rel}

As discussed in \cite{hoff_smaller_2019}, the uniformity of $p^\text{FAB}$ under $H_j$ is guaranteed regardless of the choice of prior distribution as long as the information used to compute $b_j$ is statistically independent of $T_j$. The choice of prior parameters $m_j, v_j$ must therefore be informed by quantities that are independent of the data used to compute $T_j$. In the setting of multiple biological hypotheses, such indirect information can be extracted by a probability model that describes the relationships among the biological effects $\theta_j, ~j = 1, \dots, M$. The method discussed here models these relationships as a function of historical information while using an empirical Bayes strategy to determine the relevance of that information to the current experimental data.

Specifically, let the data from a hypothesis-driven experiment be summarised by vectors $\mathbf{\bar{Y}}, \mathbf{S} \in \mathbb{R}^{M}$, where entries $\bar{Y}_{j}$ and $S_{j}$ denote the sample means and 
sample standard deviations, 
respectively, of measurements made on $\theta_j$ across $n_{j}$ replicates. Further, suppose that historical information about
each biological effect $\theta_j$ is available 
in the form of a vector of 
$q$ features derived from a large corpus of genomics data and that these vectors are collected into the rows of a matrix $\mathbf{X} \in \mathbb{R}^{M \times q}$. Then the experimental data and the biological effects can be jointly described by a sampling model for the observed data 
\begin{equation}
    \begin{aligned}
    \bar{\mathbf{Y}} | \boldsymbol{\theta} &\sim N_{M} \left(\boldsymbol{\theta}, \text{diag}(\sigma_1^2 / n_1, \dots, \sigma_M^2 / n_M) \right),  \\
    \end{aligned}
    \label{eq:sampling_model}
\end{equation} 
and a 
model relating the unobserved biological effects to the explanatory 
variables $\mathbf{X}$, 
\begin{equation}
\begin{aligned}
    \mathbf{\boldsymbol{\theta}} | \boldsymbol{\beta} &\sim N_{M} (\mathbf{X} \boldsymbol{\beta}, \tau^2 \mathbf{I}_M). \\
    \end{aligned}
    \label{eq:linking_model}
\end{equation}
Together, these two models are sometimes referred to as the 
Fay-Herriot model \citep{fay_herriot_1979,ghosh_rao_1994}, a type of hierarchical model used in the small-area estimation literature. The model (\ref{eq:linking_model}) for the parameters is referred to as the linking model, as it relates the effects of interest to historical effect-level information $\mathbf X$ 
through the parameter $\boldsymbol{\beta}$. The parameter $\tau^2$ describes 
the magnitude of the variation in the biological effects that cannot be explained by $\mathbf{X}$. The magnitude of $\boldsymbol \beta$ relative to 
$\tau^2$ is a measure of the relevance of the historical data 
to the hypothesis-driven experiment that generates 
$\bar{\mathbf{Y}}$ and $\mathbf{S}$. 

To quantify the relevance of the historical information, an empirical Bayes strategy is adopted by placing the hierarchical prior distribution
\begin{equation}
    \boldsymbol{\beta} \sim N_q \left(\mathbf{0}, \psi^2 \mathbf{I}_q\right)
\end{equation}
on the parameter $\boldsymbol{\beta}$ from the linking model (\ref{eq:linking_model}). If the variance parameters $\sigma^2, \tau^2$ and $\psi^2$ are known, then the induced conditional distribution of $\theta_j$ given all of the experimental data \textit{not} corresponding to $H_j$ is normal and can be written in closed form. The mean and variance of this distribution $\theta_j | \bar{\mathbf{Y}}_{-j}$ are then natural choices for the prior parameters $m_j$ and $v_j$. The form of these quantities is given by the following expressions
\begin{equation}\label{eq:prior_params}
    \tilde{m}_j = \mathbf{X}_j^\top \mathbf{G}_{-j}^{-1} \mathbf{X}_{-j}^\top \bar{\mathbf{Y}}_{-j}~~~~~\tilde{v}_j = \left(\frac{W_{jj}}{W_{jj} + \tau^2}\right)^2\mathbf{X}_j^\top \mathbf{G}_{-j}^{-1}\mathbf{X}_j + \tau^2.
\end{equation}
where $\mathbf{G}_{-j} = \mathbf{X}_{-j}^\top\mathbf{H}_{-j}^{-1}\mathbf{X}_{-j} + (1/\psi^{2})\mathbf{I}_q$, $\mathbf{H} = \tau^2 \mathbf{I}_M + \mathbf{W}$, $\mathbf{W} = \text{diag}(\sigma_1^2 / n_1, \dots, \sigma_M^2 / n_M)$, and $\mathbf{X}_j$ denotes the column vector containing the entries of the $j^\text{th}$ row of $\mathbf{X}$. Given a suitable estimator $\tilde{\sigma}_j^2$ that is independent of $\hat{\sigma}_j^2$, data-derived plug-in estimators for $\tau^2, \psi^2$ are obtained by maximizing the marginal likelihood of the data under a pooled variance approximation (see Section 1.1 of the Appendix). From (\ref{eq:prior_params}) it is clear that when the historical information is not relevant to the experimental data---i.e. $\tau^2 / \psi^2$ is large---prior uncertainty about $\theta_j$ is correspondingly large, and the prior mean of $\theta_j$ is close to zero. Conversely, when $\tau^2 / \psi^2$ is small, the prior mean and variance of $\theta_j$ are close to those implied by the least squares estimator of $\boldsymbol{\beta}$ under the weighted regression model $\bar{\mathbf{Y}} \sim N_M (\mathbf{X} \boldsymbol{\beta}, \mathbf{W})$, in which the biological effects are exactly a linear combination of the columns of $\mathbf{X}$.

The procedure above ensures that the prior distribution on each biological effect is strong only when the historical information is relevant to the current experimental data. Hence, the quantity
\begin{equation}
    b_j^\text{FAB} = 2 \tilde{m}_j \tilde{\sigma}_j / \sqrt{n_j} \tilde{v}_j
\end{equation}
tends to have a large absolute value only when the relevance of the historical information is high. Since $b_j^\text{FAB}$ is statistically independent of $T_j$, $H_j$ can be evaluated with $p_j^\text{FAB}$ by setting $b_j := b_j^\text{FAB}$ in (\ref{eq:p_fab}). This yields a FAB test that has the same power as the classical test if the prior information is irrelevant---i.e. as $b_j^\text{FAB} \rightarrow 0$. However, the FAB test is more powerful than the classical test if the prior information is accurate. Importantly, though, the FAB $p$-value is uniformly distributed under the null hypothesis regardless of the form or accuracy of the linking model in (\ref{eq:linking_model}). The validity of the FAB test, like that of the classical test, relies only on the validity of the sampling model in (\ref{eq:sampling_model}).

\subsection{Distilled historical information}\label{dpi}


The FAB testing procedure described above relies upon historical information pertaining to biological effects $\theta_j$ for $j = 1, \dots, M$. Because of the utility of cancer cell lines as model organisms for the study of human genomics, a vast body of genomics studies---diverse in terms of methodology, technology, and motivating research question---concern biological effects of the type $\theta_j = \theta_{lgk}$, where $l$ corresponds to a particular cancer cell line, $g$ corresponds to a particular gene, and $k$ corresponds to a genomics modality. Given a matrix $\mathbf{X}$ of historical features describing associations among cancer cell lines and among genes, the FAB testing procedure would be applicable to any study that produces measurements involving these entities. However, if the information in $\mathbf{X}$ is to be relevant to many future studies, the relationships captured by the historical features must generalize beyond any one particular modality.

Cancer genomics data corpora represent estimates of biological effects $\theta_{lgk}$ from several modalities such as RNA sequencing, DNA sequencing, CRISPR dependency, and others. Features for cancer cell lines and genes that are expressive enough to capture the structure uncovered by all of these screening modalities have the potential to be sufficiently general to be useful for FAB testing in future genomics studies. Suppose that the biological effects measured in a collection of cancer genomics data with $L$ cancer cell lines, $G$ genes, and $K$ modalities are represented by a three-way tensor $\boldsymbol{\Theta}$ with entries $\theta_{lgk}$. Then a low rank tensor approximation to $\boldsymbol{\Theta}$ can be made, so that
\begin{equation}
\theta_{lgk} \approx \mu_k +  \mathbf{U}_l^\top \mathbf{B}_k \mathbf{V}_g
\end{equation}
where $\mathbf{U}_l \in \mathbb{R}^{d_U}, \mathbf{V}_g \in \mathbb{R}^{d_V}, \mathbf{B}_k \in \mathbb{R}^{d_U \times d_V}$, and $d_U$ and $d_V$ are much smaller than $L$ and $G$, respectively. The approximation above forms the basis of a statistical model
\begin{equation}\label{eq:tens_norm}
    \begin{aligned}
		\theta_{lgk} | \mathbf{B}_k, \mathbf{U}_l, \mathbf{V}_g, \mu_k, \tau_k^2 \sim N \left( \mu_k +  \mathbf{U}_l^\top \mathbf{B}_k \mathbf{V}_g, \tau_k^2 \right)
	\end{aligned}
\end{equation}
in which the model parameters consist of a matrix $\mathbf{U}$ containing $d_U$-dimensional vector representations of cancer cell lines, a matrix $\mathbf{V}$ containing $d_V$-dimensional vector representations of genes, and a tensor $\mathbf{B}$, which modulates the inner product between these representations. In contrast to a modality-wise tensor SVD model, in which $\boldsymbol{\Theta}_k \approx \mathbf{U}_k \mathbf{B}_k \mathbf{V}_k^\top$ for each modality slice, the features in $\mathbf{U}$ and $\mathbf{V}$ are global, i.e. constant across experimental modalities. This form allows information to be shared across the different $\boldsymbol{\Theta}_k$ so that $\mathbf{U}$ and $\mathbf{V}$ encode information from all experimental modalities, while each matrix slice $\mathbf{B}_k$ selects the features of $\mathbf{U}$ and $\mathbf{V}$ local to modality $k$.

Authors \cite{ye_generalized_2005}, \cite{sutskever_modelling_2009}, and \cite{calders_bayesian_2014} consider similar models for tensor data. The probabilistic framework proposed here extends their approaches to accommodate characteristics typical of genomics datasets. For instance, for many such datasets, it is desirable to work with data values $Y_{lgk}$ that reflect an estimate of $\theta_{lgk}$ after several processing and cleaning steps have been applied to raw measurement values. For continuous-valued $Y_{lgk}$, the normal model (\ref{eq:tens_norm}) can be used by setting $Y_{lgk} = \theta_{lgk}$, so that marginally
\begin{equation}
    \begin{aligned}
		Y_{lgk} | \mathbf{B}_k, \mathbf{U}_l, \mathbf{V}_g, \mu_k, \tau_k^2 \sim N \left( \mu_k +  \mathbf{U}_l^\top \mathbf{B}_k \mathbf{V}_g, \tau_k^2 \right)
	\end{aligned}
\end{equation}
If instead the data corresponding to effects $\boldsymbol{\Theta}_k$ are binary mutation calls, the latent variable model of \citet{albert_bayesian_1993} can be adapted to the tensor model, such that marginally over $\theta_{lgk}$ the tensor data entries $Y_{lgk}$ follow a Bernoulli distribution with success probability depending on the variables $\mathbf{U}_l$, $\mathbf{V}_g$, and $\mathbf{B}_k$:
\begin{equation}\label{eq:tens_probit}
	\begin{aligned}
		&Y_{lgk} = \left\{\begin{array}{cc}
		1, & \quad \text{if} ~~ \theta_{lgk} > 0 \\ 
		0, & \quad \text{if} ~~ \theta_{lgk} < 0
		\end{array} \right. \\
		&\theta_{lgk} | \mathbf{B}_k, \mathbf{U}_l, \mathbf{V}_g, \mu_k \sim N \left( \mu_k + \mathbf{U}_l^\top \mathbf{B}_k \mathbf{V}_g, 1 \right).	\end{aligned}
\end{equation}
For data that are strictly positive and continuous---as found in RNA sequencing datasets---the ``tobit" model \citep{chib_bayes_1992} can be used, again letting data $Y_{lgk}$ be a deterministic function of $\theta_{lgk}$:
\begin{equation}\label{eq:tens_tobit}
	\begin{aligned}
		&Y_{lgk} = \left\{\begin{array}{cc}
		\theta_{lgk}, & \quad \text{if} ~~ \theta_{lgk} > 0 \\ 
		0, & \quad \text{if} ~~ \theta_{lgk} < 0
		\end{array} \right. \\
		&\theta_{lgk} | \mathbf{B}_k, \mathbf{U}_l, \mathbf{V}_g, \mu_k, \tau_k^2 \sim N \left( \mu_k + \mathbf{U}_l^\top \mathbf{B}_k \mathbf{V}_g, \tau_k^2 \right).	\end{aligned}
\end{equation}
The probabilistic formulation of the tensor model allows inference involving all of these data types to be conducted using Markov Chain Monte Carlo (MCMC). In particular, appropriate prior distributions on the model parameters lead to an efficient Gibbs sampling algorithm for sampling from the posterior distribution of the model parameters given the tensor data. Missing entries---a common feature of large genomics datasets---can be assumed Missing at Random (MAR) and marginalized out of the model within the Gibbs steps (see Section 1.3 of the Appendix). 

Closer examination of the normal tensor model reveals a connection to the linking model in (\ref{eq:linking_model}). Recall the vectorization-Kronecker identity $\text{vec}(ABC^\top) = (C \otimes A) \text{vec}(B)$ where $\text{vec}$ is defined as the operator that stacks the columns of an $m \times n$ matrix to form an $m \cdot n \times 1$ column vector \citep{horn_topics_2008}. Ignoring the intercept term $\mu_k$ for notational convenience and applying this identity to the tensor model yields a vectorized representation:
\begin{equation}
    \text{vec}(\boldsymbol{\Theta}_k) | \mathbf{B}_k, \mathbf{U}, \mathbf{V}, \tau_k^2 \sim N_{G L} \left( (\mathbf{V} \otimes \mathbf{U}) \text{vec}(\mathbf{B}_k), \tau^2_k \mathbf{I}_{G L}\right).
    \label{eq:vec_tensor}
\end{equation}
Now suppose that the model parameters $\mathbf{U}$ and $\mathbf{V}$ have been estimated from tensor data measuring biological effects from several genomics modalities. From the point of view of an experimenter running a hypothesis-driven experiment using a new modality $k'$, the matrix $\mathbf{V} \otimes \mathbf{U}$ represents historical information about the associations among cancer cell lines and among genes. According to (\ref{eq:linking_model}), the linking model for biological effects measured by the new modality is
\begin{equation}
    \boldsymbol{\theta}_{k'} | \boldsymbol{\beta}_{k'} \sim N_{G L}(\mathbf{X} \boldsymbol{\beta}_{k'}, \tau_{k'}^2 \mathbf{I}_{G L}).
\end{equation}
Replacing $\mathbf{X}$ with $\mathbf{V} \otimes \mathbf{U}$, $\boldsymbol{\beta}_{k'}$ with $\text{vec}(\mathbf{B}_{k'})$, and $\boldsymbol{\theta}_{k'}$ with $\text{vec}(\boldsymbol{\Theta}_{k'})$ in the above reveals that the linking model (\ref{eq:linking_model}) is just a special case of (\ref{eq:vec_tensor}) for a new modality $k'$. 

\begin{figure}[!h]
\centering\includegraphics[scale=0.45]{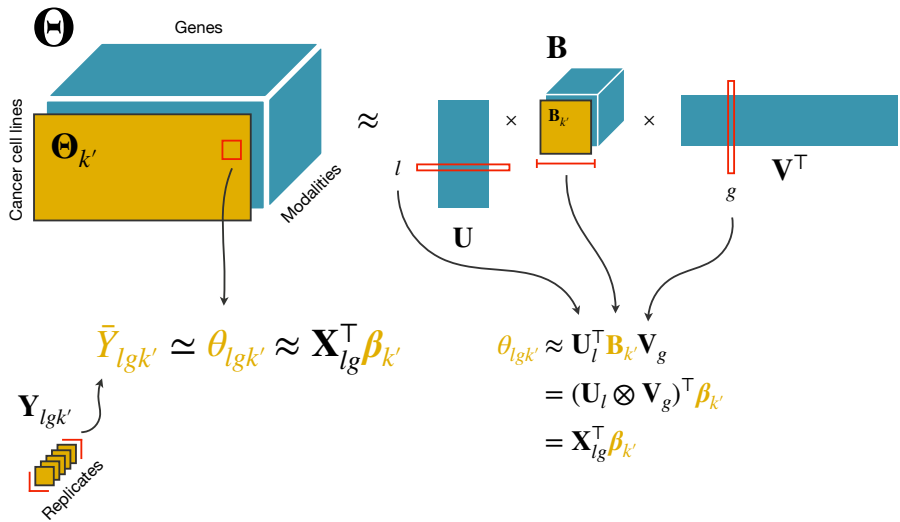}
\caption{Schematic of the FAB testing procedure. Cancer cell line and gene historical information profiles distilled from the tensor $\boldsymbol{\Theta}$ are used to model biological effects from a new modality. Experimental replicates from the new modality are assumed to be iid normal samples from the biological effects.}
\label{Fig1}
\end{figure}

Hence, for experimental data from a hypothesis-driven study, FAB tests and FAB $p$-values based on this linking model can be constructed by first estimating $\mathbf{U}$ and $\mathbf{V}$ from the tensor probability model, then testing each biological hypothesis $H_j$ by setting $\mathbf{X} = \mathbf{V} \otimes \mathbf{U}$ and following the procedure in Section \ref{fabmod} using the subset of the rows of $\mathbf{X}$ that correspond to the effects $\theta_{lgk'}$ measured in the new study (Figure \ref{Fig1}). The first step is computationally intensive, but need only be run once in advance of seeing new experimental data. The second step requires a computation for each hypothesis in the new study, but can be done efficiently using linear algebraic results for leave-one-out estimation. See Sections A.1 and A.2 of the Appendix for details on this two-step procedure.

\section{Results}
\label{sec3}

The procedure described above was applied to several simulated and real datasets to validate its type I error control, its robustness to irrelevant historical information, and its ability to yield more discoveries on real genomics data. The cancer cell line and gene features derived from the tensor probability model allowed the FAB testing procedure to be applied to measurement types as diverse as CRISPR dependency scores, drug viability profiles, differential expression scores from mammary cell subpopulations, and differential CRISPR dependency scores between mutant and wild-type cells. Though the assumed sampling model in (\ref{eq:sampling_model}) applies to experimental data with heteroscedastic error structure, in the simulation studies and investigations with real genomics data, a common sampling variance was assumed, so that $\sigma_j^2 = \sigma^2$ for all $j = 1, \dots, M$. In all cases the independent estimators $\hat{\sigma}^2, \tilde{\sigma}^2$ were calculated using an appropriate data-partitioning strategy on the entries of $\mathbf{S}$ (see Section A.1 of the Appendix).

\subsection{Simulation studies}\label{simstud}

\begin{figure}[!h]
\centering\includegraphics[scale=0.5]{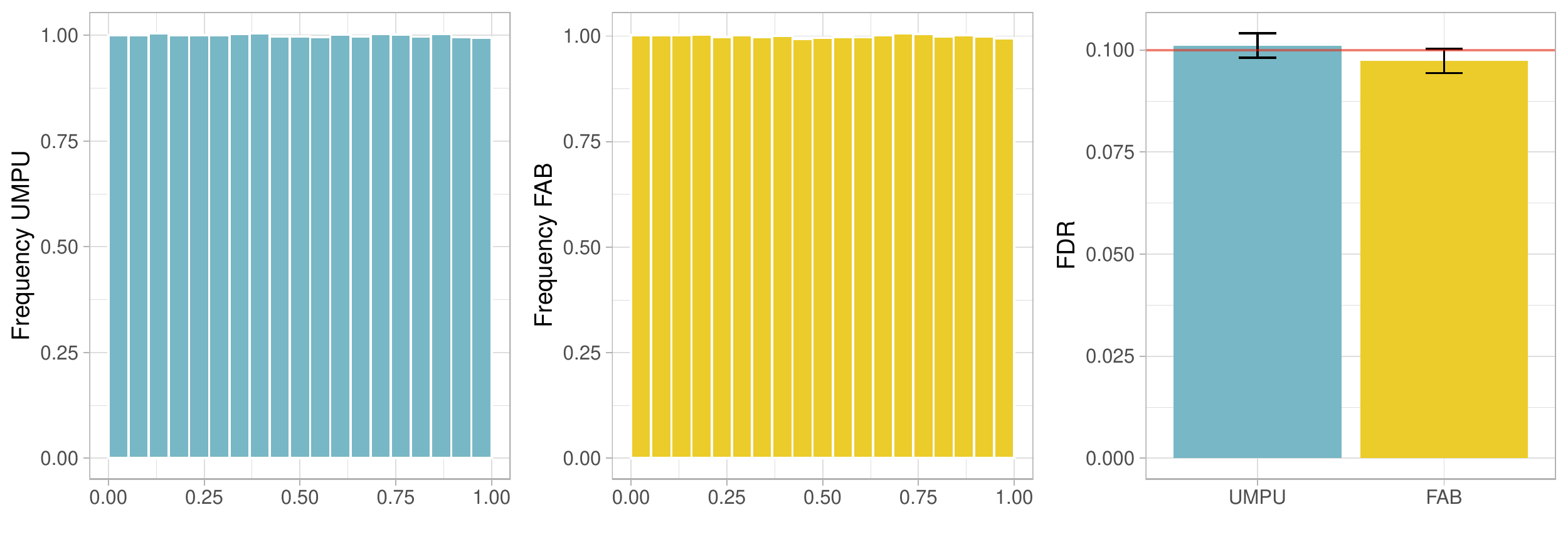}
\caption{Results of a simulation study on 10000 simulated datasets, each consisting of 10 cell lines, 25 genes, and 5 replicates. Each dataset was generated from the null sampling model ($\boldsymbol{\theta} = \boldsymbol{0}$) with $\sigma^2 = 1$. The left plot shows the distribution of $p$-values from a two-sided $t$-test. The middle plot shows the distribution of FAB $p$-values. The right plot shows the Monte Carlo average of the FDR with errorbars giving $\pm$ 1 Monte Carlo standard error. Both procedures achieve the target FDR of 0.1.}
\label{Fig2}
\end{figure}

To verify that the FAB procedure produces uniformly distributed $p$-values under the null hypothesis, $10,000$ datasets were simulated, each consisting of vectors $\mathbf{\bar{Y}}, \mathbf{S}$ of length $250$. Each entry of $\mathbf{\bar{Y}}$ and  $\mathbf{S}$ was calculated from $5$ samples drawn from a $N(0, 1)$ distribution, representing $5$ replicates of a null measurement with sampling variance equal to 1. Matrices $\mathbf{U} \in \mathbb{R}^{10 \times 5}, \mathbf{V} \in \mathbb{R}^{25 \times 5}$, representing, respectively, cancer cell line and gene features were then generated with entries drawn from $N(0, 1)$. The distribution of FAB $p$-values obtained from fitting models (\ref{eq:sampling_model}), (\ref{eq:linking_model}) to each of the $10, 000$ datasets was empirically uniform, indistinguishable from that of the classical $p$-values (Figure \ref{Fig2}). To demonstrate proper FDR control, the $p$-values obtained for each dataset were then adjusted using the Benjamini-Hochberg (BH) procedure \citep{benjamini_controlling_1995}. Since all of the values in each of the datasets were simulated from null effects, a value of $1$ was recorded for a dataset if any of its BH-adjusted $p$-values fell below a target of $0.1$. A value of $0$ was recorded otherwise. Taking the average of the $0$'s and $1$'s over the $10,000$ trials yielded Monte Carlo estimates of the FDR for the classical and FAB tests. Both achieved the target FDR of $0.1$ up to Monte Carlo error (Figure \ref{Fig2}). 

\begin{figure}[!h]
\centering\includegraphics[scale=0.55]{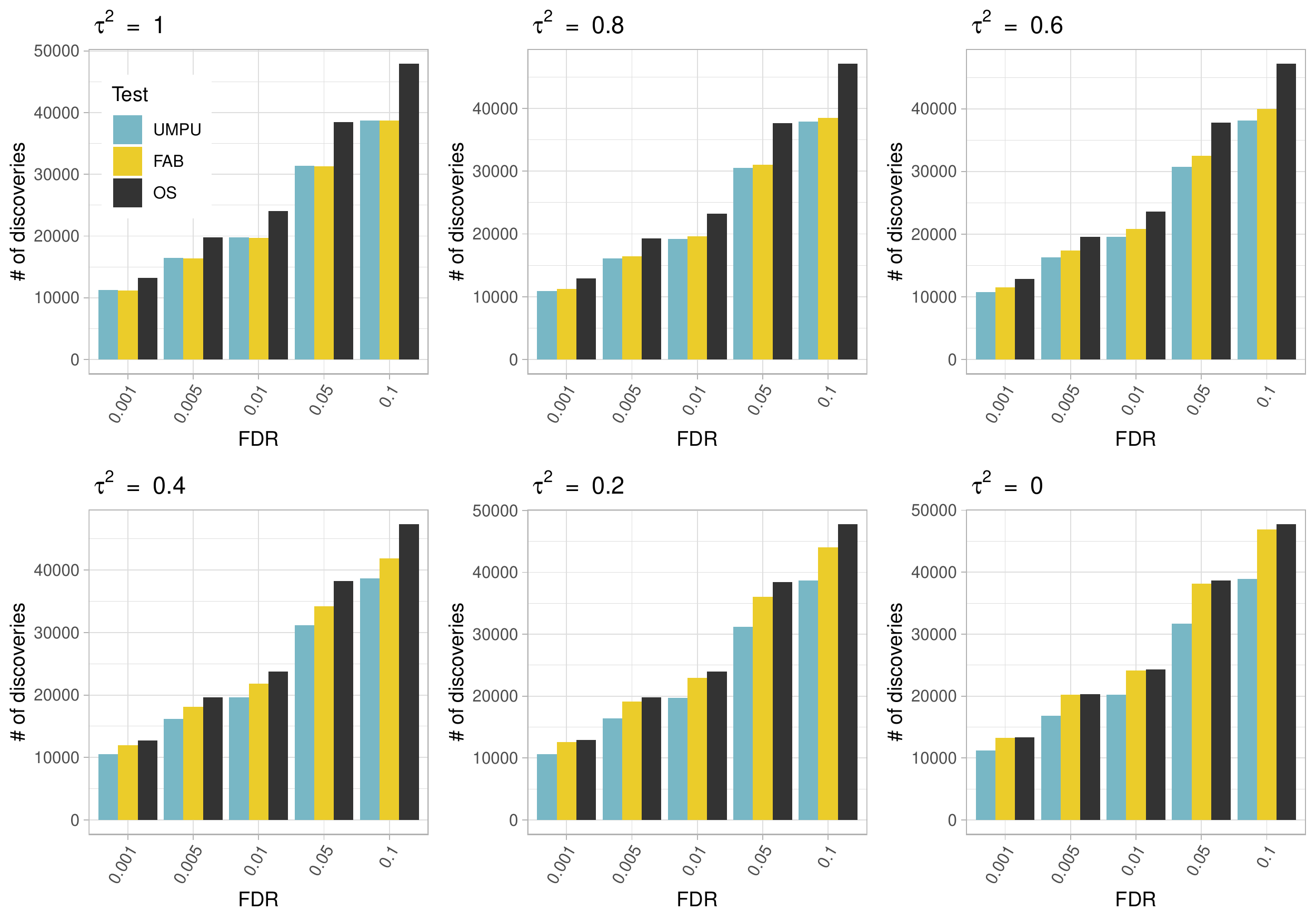}
\caption{Results of a simulation study demonstrating the effect of increasing model precision on the behavior of the FAB test. The plot titles show the proportion of variance in $\boldsymbol{\theta}$ unexplained by $\mathbf{X}$. The height of the bars represent the cumulative number of discoveries made at increasing FDR levels for two-sided classical (light blue), FAB (yellow), and one-sided oracle (dark grey) tests. When the lack-of-fit variance $\tau^2$ is close to the total variance in the biological effects, the FAB test yields as many discoveries as the standard two-sided test. When $\tau^2$ is small, the FAB test behaves more like a one-sided oracle test.}
\label{Fig3}
\end{figure}

Further simulations were conducted to test the behavior of the FAB procedure under non-null scenarios. First, matrices $\mathbf{{U}} \in \mathbb{R}^{5 \times 10}, \mathbf{{V}} \in \mathbb{R}^{100 \times 10}$ were generated with entries drawn from $N(0, 1)$ and a matrix $\mathbf{X}$  was set to $\mathbf{X} := \mathbf{V} \otimes \mathbf{U}$. Then, vectors $\boldsymbol{\beta} \in \mathbb{R}^{100}, \boldsymbol{\epsilon} \in \mathbb{R}^{500}$ were generated, each with entries drawn from $N(0, 1)$. The vectors $\mathbf{X} \boldsymbol{\beta}$ and $\boldsymbol{\epsilon}$ were then normalized to have empirical variance equal to 1. For $\tau^2$---the magnitude of variation left unexplained by $\mathbf{X}$---taking values in $\{1, 0.8, 0.6, 0.4, 0.2, 0\}$, a biological effects vector was constructed according to
\begin{equation*}
    \boldsymbol{\theta} = \sqrt{1-\tau^2} \mathbf{X}\boldsymbol{\beta} + \sqrt{\tau^2}  \boldsymbol{\epsilon}
\end{equation*}
and $200$ datasets $\bar{\mathbf{Y}}, \mathbf{S}$ were simulated for each value of $\tau^2$ according to $Y_{ij} \sim N(\theta_j, 1),~i = 1, \dots, 5$. The simulation procedure therefore ensured that the total variation in $\boldsymbol{\theta}$ was held constant at $1$, while the proportion of variation explained by $\mathbf{X}$ increased as $\tau^2$ decreased. The cumulative number of discoveries made across the 200 trials was then recorded at varying FDR thresholds. As shown in Figure \ref{Fig3}, when $\tau^2 = 1$, the FAB procedure yielded as many discoveries as the classical two-sided test. However, under all other conditions, the FAB test produced more discoveries at every FDR threshold. As $\tau^2$ decreased the number of discoveries made by the FAB test cleanly interpolated between that of the classical two-sided test and that of a one-sided oracle test against the alternative in the direction of $\text{sign}(\theta_j)$.

\subsection{Smaller $p$-values in genomics studies} \label{examples}

To extract biologically meaningful historical information, the tensor factorization method described in Section \ref{dpi} was applied to a collection of four genomics datasets downloaded from the Cancer Dependency Map portal \citep{broad_depmap_depmap_2019}. Each dataset contains the results of a different screening technology used to measure genetic attributes of a set of cancer cell lines. The gene expression dataset \citep{barretina_cancer_2012} contains measurements of mRNA abundance, which is a proxy for the amount of the gene-product active in the cell. The mutation dataset \citep{barretina_cancer_2012} contains binary entries indicating whether gene $g$ is mutated in cell line $l$. The CRISPR \citep{meyers_computational_2017} and RNAi \citep{mcfarland_improved_2018} datasets each contain cell viability scores corresponding to the growth inhibitory effect induced in cell line $l$ when gene $g$ is deleted (CRISPR) or suppressed (RNAi).

The tensor probability model was fit to these four modalities using the tobit model (\ref{eq:tens_tobit}) for the gene expression dataset, the probit model (\ref{eq:tens_probit}) for the mutation dataset, and the normal model (\ref{eq:tens_norm}) for the CRISPR and RNAi datasets. Historical information profiles of dimension $d_U = 16$ and $d_V = 64$ were distilled for 1209 cancer cell lines and 4570 genes, respectively (Section A.5 of the Appendix). Appropriate prior distributions on the tensor model parameters yielded closed form conditional probability densities, which were used to obtain $5,000$ samples from the posterior distributions of all model parameters via a Gibbs sampling algorithm (Section A.3 of the Appendix). The first $1,000$ samples were discarded, and estimators of the cancer cell line and gene historical information profiles were obtained by taking the average of the remaining $4,000$ posterior samples of $\mathbf{U}$ and $\mathbf{V}$ after aligning them with a Procrustes procedure (Section A.4 of the Appendix). Subsets of the matrix $\mathbf{X} = \mathbf{V} \otimes \mathbf{U}$ were then used as predictors in the linking model (\ref{eq:linking_model}) to perform FAB tests for hypotheses in several genomics studies.

\subsubsection{CRISPR essentiality screens in human AML cell lines}

\begin{figure}[!h]
\centering\includegraphics[scale=0.7]{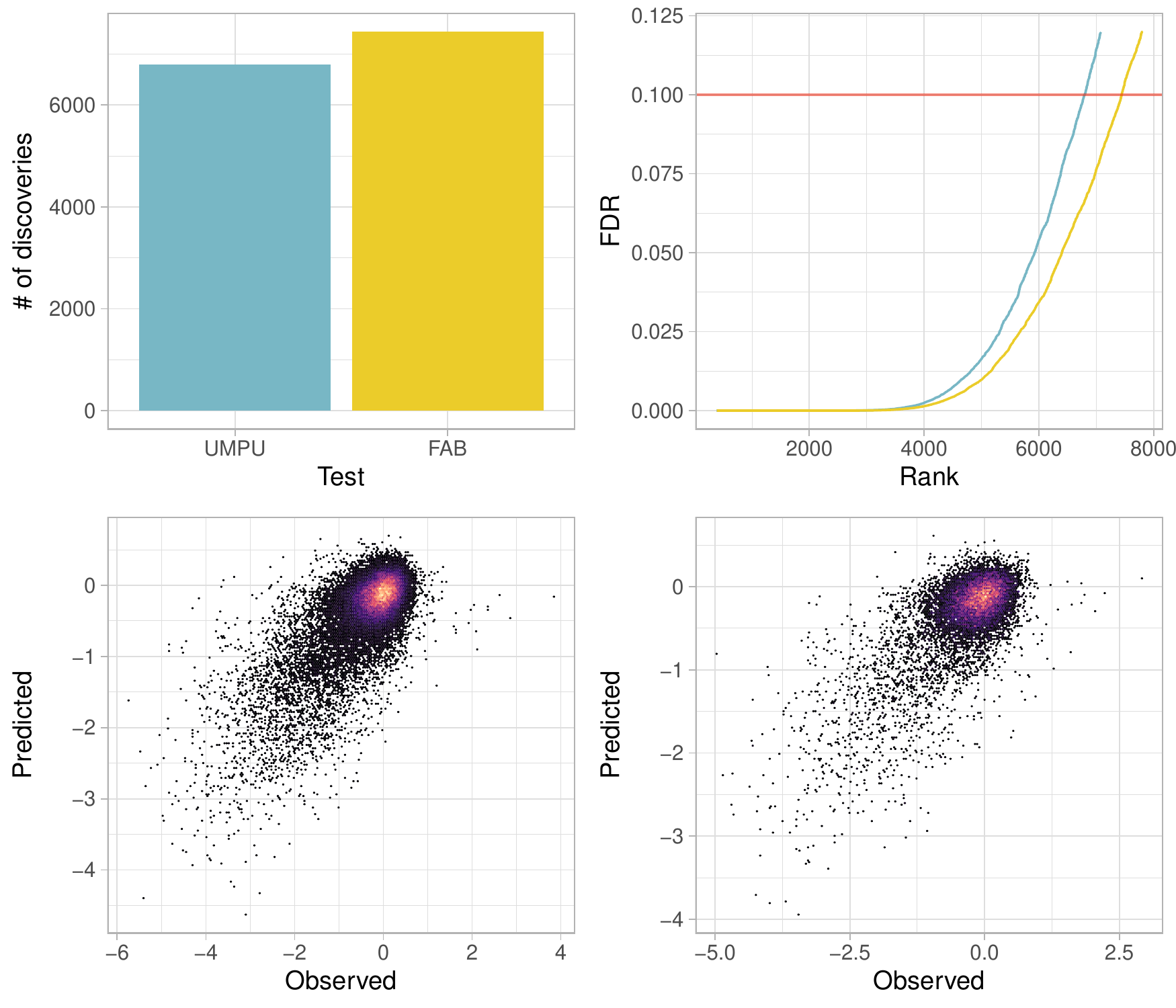}
\caption{Results of applying the linear model to data from the \cite{wang_gene_2017} CRISPR essentiality screens. In the upper left plot, the number of discoveries at an FDR of $0.1$ is plotted for each test.  In the upper right plot, BH-adjusted $p$-values for standard two-sided and FAB tests are plotted against the rank of the adjusted $p$-value. The FAB test leads to more discoveries for a range of FDRs. The lower plots show predicted CRISPR dependency scores against observed CRISPR dependency scores for cell lines that also appear in the DepMap CRISPR dataset (left) and those that do not (right).}
\label{Fig4}
\end{figure}

To demonstrate the utility of the FAB testing framework on genomic screening modalities already contained in the Dependency Map collection, dependency scores for $11$ acute myeloid leukemia (AML) cancer cell lines and $3537$ genes were obtained from genome-wide CRISPR screens conducted by \cite{wang_gene_2017}. The results of applying the FAB testing procedure using the corresponding cancer cell line and gene historical features are displayed in Figure \ref{Fig4}. As shown, the FAB test produced more discoveries than the standard two-sided test at a range of FDR levels. The observed increase in power was a result of the relatively high degree of predictive model fit (Figure \ref{Fig4}, bottom row). Many of the AML cell lines in the \cite{wang_gene_2017} dataset were also screened with the CRISPR technology by \cite{broad_depmap_depmap_2019}; hence, it is reassuring though perhaps not surprising that the model had good predictive performance on the data corresponding to these cell lines (Figure \ref{Fig4}, bottom left). Of note is the fact that the model also showed good predictive performance on the $3$ cell lines \textit{not} screened as part of the Dependency Map CRISPR effort (Figure \ref{Fig4}, bottom right). For these cell lines, the information pertaining to CRISPR dependency was truly indirect, coming from other cancer cell lines and other modalities in the Dependency Map collection.

\subsubsection{Viability profiling to repurpose non-oncology drugs}

\begin{figure}[!h]
\centering\includegraphics[scale=0.6]{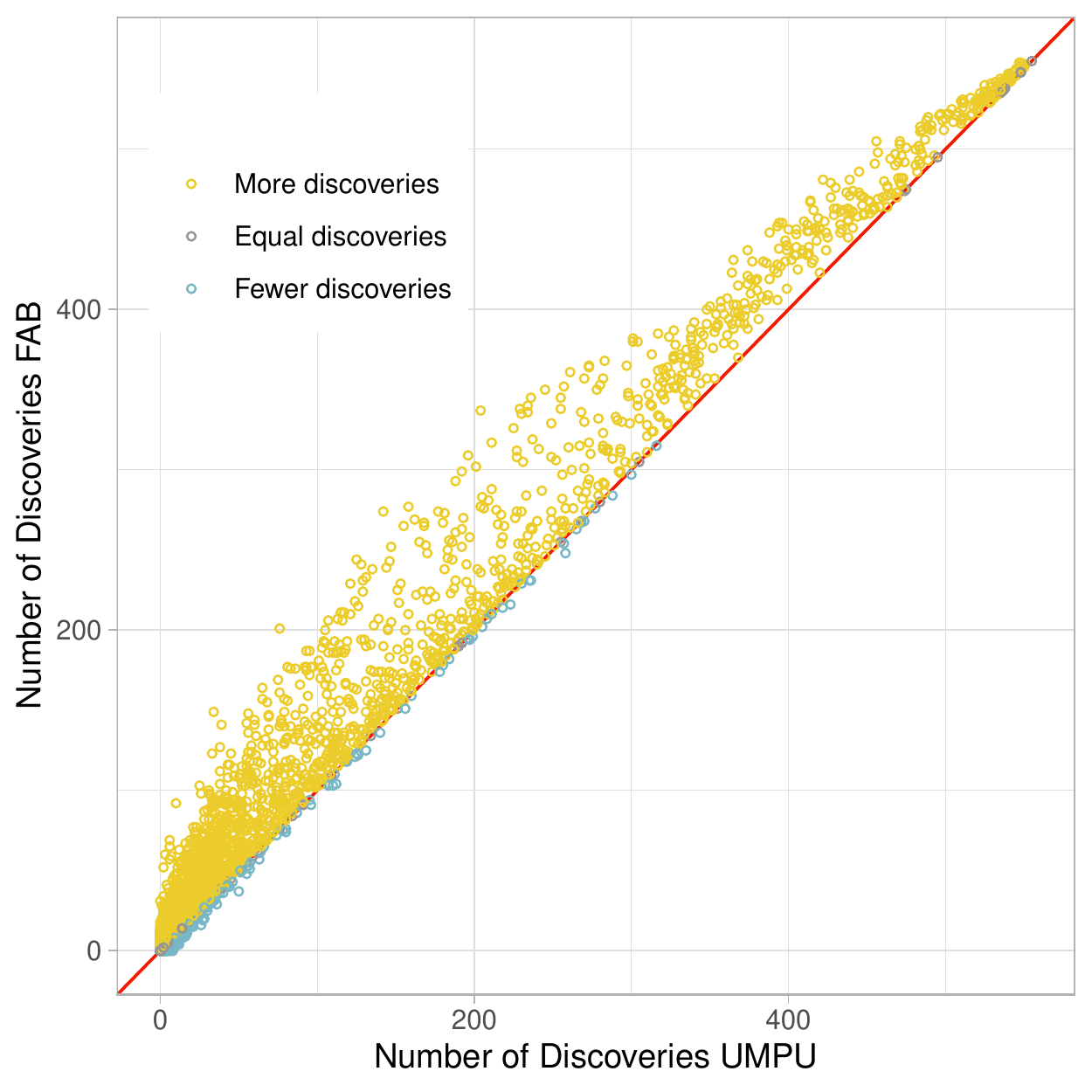}
\caption{Number of discoveries made at an FDR of 0.1 for each compound profile in Repurposing dataset for classical and FAB $p$-values. Drugs for which more discoveries were made by the FAB test are highlighted in yellow, while drugs for which more discoveries were made by the classical test are highlighted in blue. Drugs for which the number of discoveries was equal are highlighted grey and lie along the red diagonal line.}
\label{Fig5}
\end{figure}

The use of historical information distilled from the Dependency Map need not be limited to experimental designs with many cell lines and many genes. The FAB testing procedure is also applicable to a class of experiments known as ``cell viability profiling" in which a perturbagen, such as a drug, is applied to a panel of cell lines in order to determine which cell lines are sensitive or resistant to the perturbagen's action. Negative viability scores correspond to cell-death or growth-inhibition in response to the drug treatment, while positive scores suggest enhanced proliferation. A recent effort to uncover oncology indications for non-oncology drugs \citep{corsello_non-oncology_2019} produced $4686$ such viability profiles, each corresponding to a drug's action on a panel of $578$ cancer cell lines, $569$ of which appear in the Dependency Map dataset. For these experiments, the cell line features were used to model each of the drug viability profiles. Using the notation from Section \ref{dpi}, the procedure was equivalent to setting $\mathbf{V} := 1$. As shown in Figure \ref{Fig5}, at an FDR of $0.1$, the FAB tests yielded as many or more discoveries than traditional $p$-values for $80$\% of the drugs screened. For $67$\% of the drugs screened, the FAB test yielded strictly more discoveries. As measured by absolute deviance from the red line on the main diagonal of Figure \ref{Fig5}, the difference between the number of discoveries made by the FAB and classical tests tended to be small when the FAB test yielded fewer discoveries and tended to be larger when the FAB test yielded more discoveries.

\subsubsection{Differential expression analysis for mammary cell subpopulations}

Experimental data containing measurements for hundreds of cancer cell lines are relatively scarce due to the labour-intensive nature of cell culture. More common in the genomics literature are datasets with treatment-control measurements for a limited number of samples, but many genes. These data are often interpreted by performing a differential expression analysis, in which significant genes are identified by conducting hypothesis tests on contrast scores between samples in the treatment and control conditions for each gene in a study. 

To demonstrate the utility of the FAB procedure in differential expression analysis, gene expression profiles of mammary cell subpopulations from three healthy human donors were obtained from a study conducted by \cite{lim_aberrant_2009}. Each donor in the study contributed samples of mammary stem cells (MS), luminal progenitor cells (LP) and mature luminal cells (ML). The object of the study was to test for significantly differentially expressed genes between MS / LP, MS / ML, and ML / LP cell populations. To prepare the data for use, the steps outlined in the tutorial for the software package \textit{limma} \citep{ritchie_limma_2015, ritchie_comparison_2007, smyth_normalization_2003} were followed. For each contrast, the FAB testing model was applied to the average difference in gene expression between the cell populations, using the gene features $\mathbf{V}$.

\begin{figure}[!h]
\centering\includegraphics[scale=0.4]{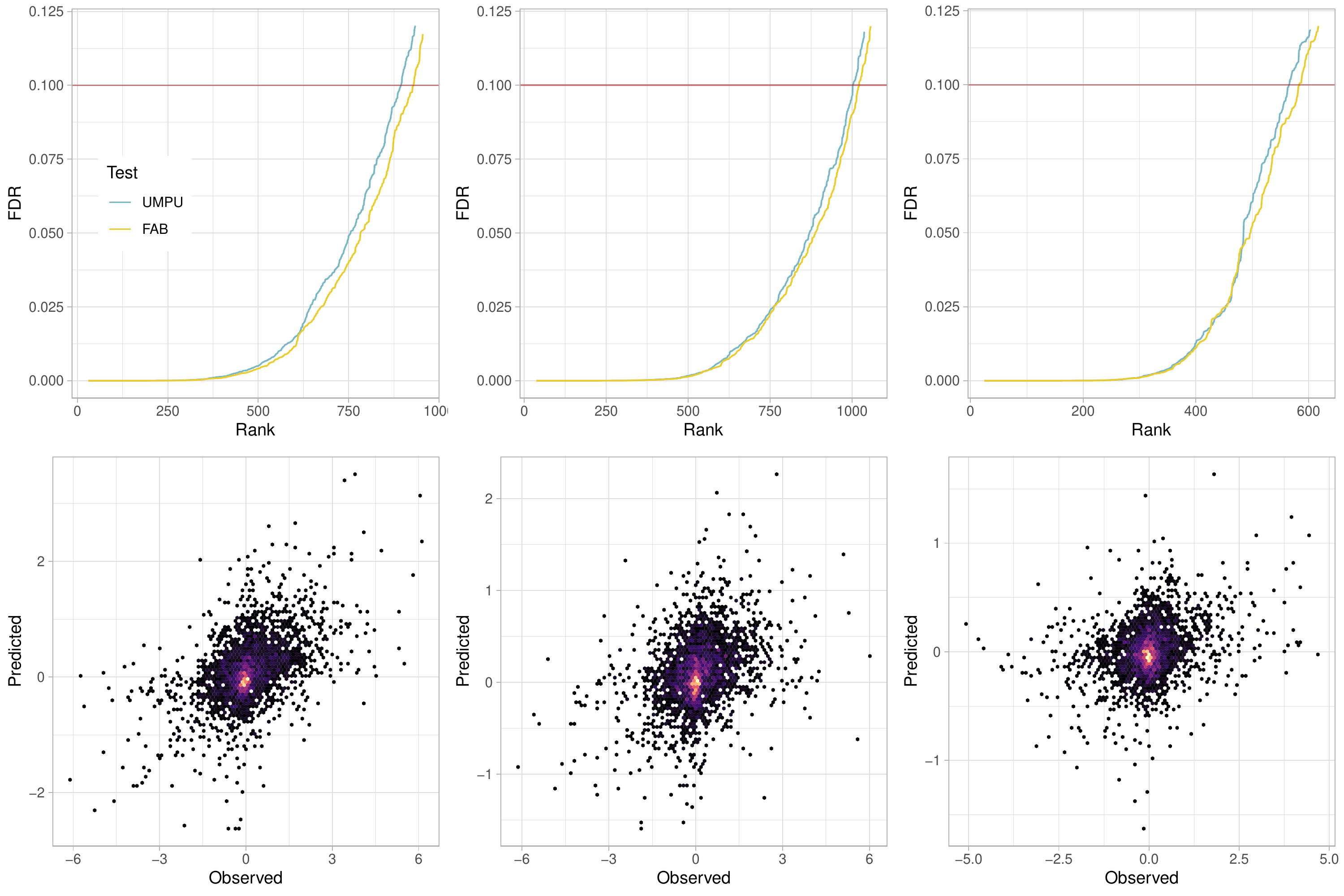}
\caption{Results of applying the linear model to data from \cite{lim_aberrant_2009}. From left to right, the contrasts tested were MS vs. LP, MS vs. ML, and ML vs. LP. Although the model precision is far from perfect (bottom row), the FAB test is more powerful at an FDR of $0.1$ for all three contrasts.}
\label{Fig6}
\end{figure}

Although the linking model precision varied across the different contrasts, it showed some capacity for predicting all of the differential expression scores (Figure \ref{Fig6}). Correspondingly, the FAB $p$-values yielded more discoveries at an FDR of $0.1$ for each of the three contrasts than those from a traditional two-sided $t$-test. The fact that the FAB approach was able to achieve some predictive power in this setting is notable for two important reasons: first, it suggests that the FAB framework can be applied to modalities---such as contrast scores---that bear very little relationship to those contained in the dataset from which the historical information was distilled. Second, it suggests that the gene-to-gene relationships distilled by the tensor model from a corpus of cancer genomics data generalize to non-cancer biological contexts.

\subsubsection{Robustness of the FAB test}

The FAB testing procedure convincingly yields more discoveries than the classical two-sided test when applied to experimental data with some degree of linear relationship to historical features. To illustrate what happens in practice when there is little to no linear relationship between the historical features and the experimental data, CRISPR dependency scores from the \cite{kory_sfxn1_2018} study were obtained for 2 cancer cell lines and $2865$ genes. The dependency scores consist of measurements for each cell line under two experimental conditions: culture in media that contained the $\alpha$-amino acid serine, and culture in serine-free media. Differential dependency scores were calculated by taking the difference in average dependency score between the culture conditions.

\begin{figure}[!h]
\centering\includegraphics[scale=0.35]{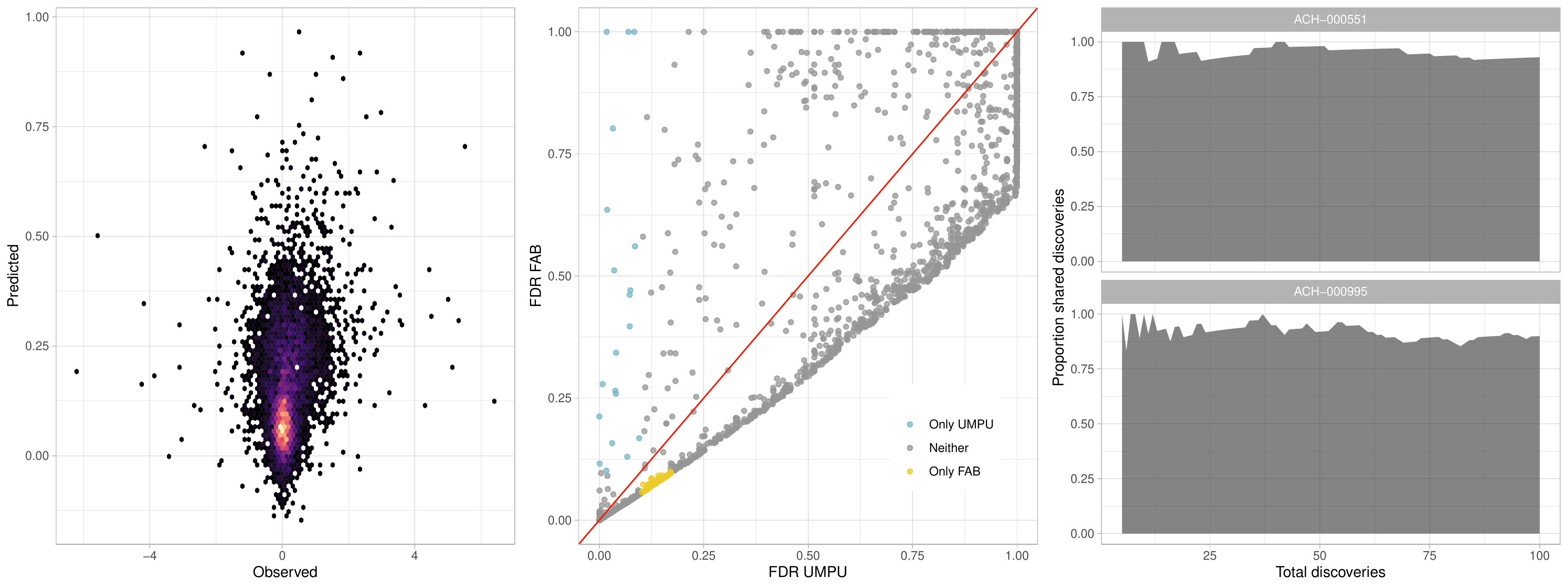}
\caption{Results of applying the linear model to data from the \cite{kory_sfxn1_2018} study. The model predictions are uncorrelated with the observed differential dependency scores (left). Still, at an FDR of 0.1 relatively few discoveries are made only by the classical test (center, blue), while relatively many are made only by the FAB test (center, yellow). Moreover, the list of top 1-100 most significant genes within each cancer cell line for the FAB and classical tests are nearly identical to one another (right).}
\label{Fig7}
\end{figure}

In contrast to previous examples, the differential dependency scores predicted by the linking model showed almost no correlation to the observed data from this study (Figure \ref{Fig7}, left). The FAB test yielded more discoveries than the classical test, but the increase in power was due to the fact that most of the observed differential dependency scores had positive values. Hence the predicted values from the linking model were centered about a positive intercept term. However, the scale of the predicted values was appropriately small relative to that of the observed data due to the shrinkage induced by the empirical Bayes estimator $\tilde{\psi}^2$. Likewise, the estimator $\tilde{\tau}^2$ was appropriately large, so that the computed offset terms $b^{\text{FAB}}_j$ were conservative relative to the classical test statistics $T_j$. The result was that, at an FDR of $0.1$, relatively few discoveries were made only by the classical test, while relatively many discoveries were made only by the FAB test (Figure \ref{Fig7}, middle). For both cancer cell lines in the study, the lists of genes produced by taking, respectively, the $r$-smallest adjusted classical and FAB $p$-values also showed a high degree of overlap for ranks $r = 1, \dots, 100$ (Figure \ref{Fig7}, right).

\subsubsection{Biological relevance of FAB discoveries}

\begin{figure}[!h]
\centering\includegraphics[scale=0.4]{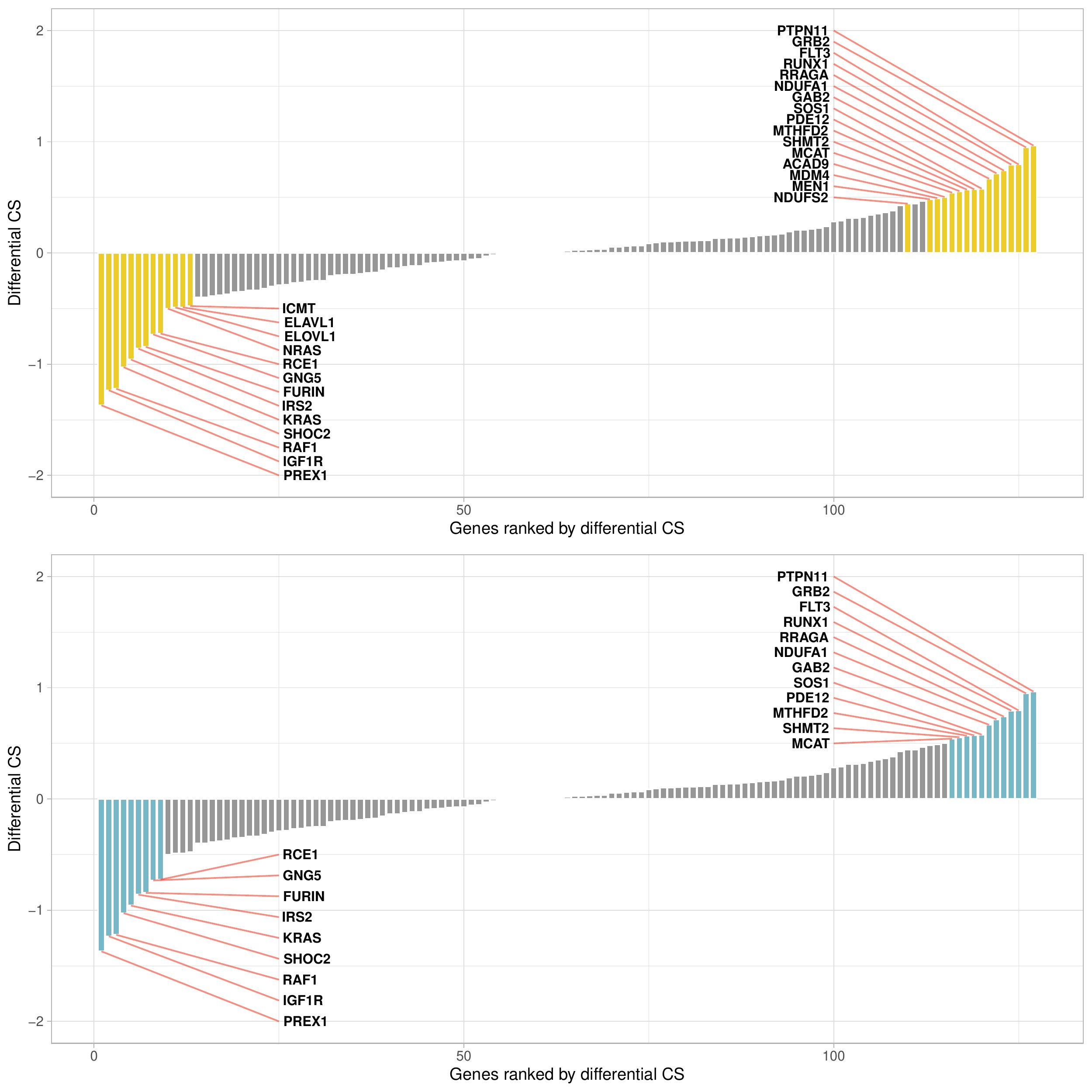}
\caption{Results of applying the linear model to data from the follow-up study in \cite{wang_gene_2017}. The genes discovered at a 0.1 FDR threshold are displayed above for the FAB test and below for the classical test.}
\label{Fig8}
\end{figure}

While the FAB testing procedure has the potential to increase the expected number of discoveries made in a given study, it does not give any guarantees about the quality of those discoveries. The latter is difficult to analyze in the abstract. In practice, however, the quality of additional discoveries in genomics studies may be analyzed by evaluating their biological plausibility. To this end, FAB tests were performed in a differential dependency analysis between $10$ Ras-mutant and $8$ Ras-wild-type AML cell lines using data from a follow-up experiment conducted by \cite{wang_gene_2017}. The follow-up experiment measured CRISPR dependency scores for a focused set of $127$ genes. Some of the genes were previously suspected to be interactors with oncogenic Ras and others were included as controls. The discoveries made by the FAB and classical tests at an FDR of $0.1$ are highlighted in Figure \ref{Fig8}.

Overall, the FAB test yielded $30$ discoveries, compared to $23$ discoveries from the classical test. Among the genes discovered only by the FAB test was \textit{NRAS}, a member of the Ras family and one of the genes used to distinguish Ras-mutants from wild-type. Ras-mutant cancer cell lines tend to show increased dependency in Ras family genes (note that the oncogene \textit{KRAS} is one of the primary discoveries made by both tests) compared to wild-type cancer cell lines, so the inclusion of \textit{NRAS} among the list of discoveries is biologically plausible. Another discovered gene, \textit{ICMT}, was a hit in a separate experiment from the \cite{wang_gene_2017} article and is known to be involved in the maturation of Ras. Yet another gene discovered by the FAB test alone, \textit{ELAVL1}, is a putative regulator of cell survival in KRAS-driven pancreatic ductal adenocarcinomas (PDACs) \citep{brody_complex_2018}, though its role in the context of AML is not as well understood. Finally, the gene \textit{ELOVL1}, discovered only by the FAB test, is a fatty acid elongase that performs cell-essential fatty acid synthesis. It has been reported that Ras-driven tumors activate fatty acid synthesis genes, and that this activation creates a synthetic lethality with Rapamycin \citep{gouw_oncogene_2017, salloum_mutant_2014}. Further, \cite{guo_rapamycin_2016} note that Rapamycin supresses expression of \textit{ELOVL1} in bovine mammary epithelial cells (BMECs). Hence, loss-of-function of \textit{ELOVL1} could plausibly have a differentially adverse effect on Ras-mutant cancer cell lines like those in the \cite{wang_gene_2017} follow-up study. Together, these results suggest that the additional genes discovered by the FAB test are likely from the set of biologically relevant Ras-interactors, rather than from the set of inactive controls.

\section{Discussion}
\label{sec4}

Given a sufficiently rich source of historical information, hypothesis tests in genomics studies can realize substantial gains in statistical power under the FAB testing framework while still maintaining experiment-wise type I error rates. A probability model that describes relationships among biological effects can be used not only to synthesize this source of indirect information, but also to construct a distribution according to which FAB $p$-values are computed. Regardless of whether this probability model is correct, however, the resulting FAB $p$-values are uniformly distributed under the null hypothesis. In the case of model misspecification, parameters that inform the degree to which historical information is relevant to a given genomics study can be estimated via an empirical Bayes strategy in order to protect against over-confidence in inferred quantities used in the FAB tests. Both in simulation and in practice, this procedure adaptively reverts FAB tests to classical tests, so that bona fide discoveries are not obscured when the model fit is poor. This suggests that the FAB testing framework can safely be applied even when the relevance of the historical information to data from a new study is unknown. Practical investigations demonstrate that historical features distilled from the Cancer Dependency Map dataset by the tensor probability model contain information relevant to data types ranging from direct readouts of familiar screening technologies to esoteric contrast scores between distinct, non-cancerous biological populations. A promising research direction is to consider whether there are other ways in which these features might be used to inform practitioners of hypothesis-driven genomics experiments. 

\section{Software}
\label{sec5}

Software in the form of R code is available at \url{https://github.com/j-g-b/BTF}.

\section*{Acknowledgments}

The authors thank Charles Henry Adelmann for helpful conversations.

\bibliographystyle{biorefs}
\bibliography{refs}

\begin{thebibliography}{99}

\bibitem[Albert and Chib(1993)Albert and Chib]{albert_bayesian_1993}
\textsc{Albert, James~H. and Chib, Siddhartha}. (1993, June).
\newblock Bayesian {Analysis} of {Binary} and {Polychotomous} {Response}
  {Data}.
\newblock {\em Journal of the American Statistical
  Association\/}~\textbf{88}(422), 669.

\bibitem[Barretina \emph{and others}(2012)Barretina, Caponigro, Stransky,
  Venkatesan, Margolin, Kim, Wilson, Lehár, Kryukov, Sonkin, Reddy, Liu,
  Murray, Berger, Monahan, Morais, Meltzer, Korejwa, Jané-Valbuena, Mapa,
  Thibault, Bric-Furlong, Raman, Shipway, Engels, Cheng, Yu, Yu, Aspesi,
  de~Silva, Jagtap, Jones, Wang, Hatton, Palescandolo, Gupta, Mahan, Sougnez,
  Onofrio, Liefeld, MacConaill, Winckler, Reich, Li, Mesirov, Gabriel, Getz,
  Ardlie, Chan, Myer, Weber, Porter, Warmuth, Finan, Harris, Meyerson, Golub,
  Morrissey, Sellers, Schlegel and Garraway]{barretina_cancer_2012}
\textsc{Barretina, Jordi, Caponigro, Giordano, Stransky, Nicolas, Venkatesan,
  Kavitha, Margolin, Adam~A.  \emph{and others}}. (2012, March).
\newblock The {Cancer} {Cell} {Line} {Encyclopedia} enables predictive
  modelling of anticancer drug sensitivity.
\newblock {\em Nature\/}~\textbf{483}(7391), 603--607.

\bibitem[Behan \emph{and others}(2019)Behan, Iorio, Picco, Gonçalves, Beaver,
  Migliardi, Santos, Rao, Sassi, Pinnelli, Ansari, Harper, Jackson, McRae,
  Pooley, Wilkinson, van~der Meer, Dow, Buser-Doepner, Bertotti, Trusolino,
  Stronach, Saez-Rodriguez, Yusa and Garnett]{behan_prioritization_2019}
\textsc{Behan, Fiona~M., Iorio, Francesco, Picco, Gabriele, Gonçalves,
  Emanuel, Beaver, Charlotte~M.  \emph{and others}}. (2019, April).
\newblock Prioritization of cancer therapeutic targets using {CRISPR}–{Cas9}
  screens.
\newblock {\em Nature\/}~\textbf{568}(7753), 511--516.

\bibitem[Benjamini and Hochberg(1995)Benjamini and
  Hochberg]{benjamini_controlling_1995}
\textsc{Benjamini, Yoav and Hochberg, Yosef}. (1995, January).
\newblock Controlling the {False} {Discovery} {Rate}: {A} {Practical} and
  {Powerful} {Approach} to {Multiple} {Testing}.
\newblock {\em Journal of the Royal Statistical Society: Series B
  (Methodological)\/}~\textbf{57}(1), 289--300.

\bibitem[Birsoy \emph{and others}(2015)Birsoy, Wang, Chen, Freinkman,
  Abu-Remaileh and Sabatini]{birsoy_essential_2015}
\textsc{Birsoy, Kıvanç, Wang, Tim, Chen, Walter~W., Freinkman, Elizaveta,
  Abu-Remaileh, Monther and Sabatini, David~M.} (2015, July).
\newblock An {Essential} {Role} of the {Mitochondrial} {Electron} {Transport}
  {Chain} in {Cell} {Proliferation} {Is} to {Enable} {Aspartate} {Synthesis}.
\newblock {\em Cell\/}~\textbf{162}(3), 540--551.

\bibitem[Brody and Dixon(2018)Brody and Dixon]{brody_complex_2018}
\textsc{Brody, Jonathan~R. and Dixon, Dan~A.} (2018, May).
\newblock Complex {HuR} function in pancreatic cancer cells.
\newblock {\em Wiley Interdisciplinary Reviews: RNA\/}~\textbf{9}(3), e1469.

\bibitem[Chib(1992)Chib]{chib_bayes_1992}
\textsc{Chib, Siddhartha}. (1992, January).
\newblock Bayes inference in the {Tobit} censored regression model.
\newblock {\em Journal of Econometrics\/}~\textbf{51}(1-2), 79--99.

\bibitem[Corsello \emph{and others}(2019)Corsello, Nagari, Spangler, Rossen,
  Kocak, Bryan, Humeidi, Peck, Wu, Tang, Wang, Bender, Lemire, Narayan,
  Montgomery, Ben-David, Chen, Rees, Lyons, McFarland, Wong, Wang, Dumont,
  O’Hearn, Stefan, Doench, Greulich, Meyerson, Vazquez, Subramanian, Roth,
  Bittker, Boehm, Mader, Tsherniak and Golub]{corsello_non-oncology_2019}
\textsc{Corsello, Steven~M., Nagari, Rohith~T., Spangler, Ryan~D., Rossen,
  Jordan, Kocak, Mustafa  \emph{and others}}. (2019, August).
\newblock Non-oncology drugs are a source of previously unappreciated
  anti-cancer activity.
\newblock {\em preprint}, Cancer Biology.

\bibitem[DepMap(2019)DepMap]{broad_depmap_depmap_2019}
\textsc{DepMap, Broad}. (2019).
\newblock {DepMap} {19Q3} {Public}.
\newblock type: dataset.

\bibitem[Fay and Herriot(1979)Fay and Herriot]{fay_herriot_1979}
\textsc{Fay, Robert~E. III and Herriot, Roger~A.} (1979).
\newblock Estimates of income for small places: an application of
  {J}ames-{S}tein procedures to census data.
\newblock {\em J. Amer. Statist. Assoc.\/}~\textbf{74}(366, part 1), 269--277.

\bibitem[Ghandi \emph{and others}(2019)Ghandi, Huang, Jané-Valbuena, Kryukov,
  Lo, McDonald, Barretina, Gelfand, Bielski, Li, Hu, Andreev-Drakhlin, Kim,
  Hess, Haas, Aguet, Weir, Rothberg, Paolella, Lawrence, Akbani, Lu, Tiv,
  Gokhale, de~Weck, Mansour, Oh, Shih, Hadi, Rosen, Bistline, Venkatesan,
  Reddy, Sonkin, Liu, Lehar, Korn, Porter, Jones, Golji, Caponigro, Taylor,
  Dunning, Creech, Warren, McFarland, Zamanighomi, Kauffmann, Stransky,
  Imielinski, Maruvka, Cherniack, Tsherniak, Vazquez, Jaffe, Lane, Weinstock,
  Johannessen, Morrissey, Stegmeier, Schlegel, Hahn, Getz, Mills, Boehm, Golub,
  Garraway and Sellers]{ghandi_next-generation_2019}
\textsc{Ghandi, Mahmoud, Huang, Franklin~W., Jané-Valbuena, Judit, Kryukov,
  Gregory~V., Lo, Christopher~C.  \emph{and others}}. (2019, May).
\newblock Next-generation characterization of the {Cancer} {Cell} {Line}
  {Encyclopedia}.
\newblock {\em Nature\/}~\textbf{569}(7757), 503--508.

\bibitem[Ghosh and Rao(1994)Ghosh and Rao]{ghosh_rao_1994}
\textsc{Ghosh, M. and Rao, J. N.~K.} (1994).
\newblock Small area estimation: an appraisal.
\newblock {\em Statist. Sci.\/}~\textbf{9}(1), 55--93.
\newblock With comments and a rejoinder by the authors.

\bibitem[Gouw \emph{and others}(2017)Gouw, Eberlin, Margulis, Sullivan, Toal,
  Tong, Zare and Felsher]{gouw_oncogene_2017}
\textsc{Gouw, Arvin~M., Eberlin, Livia~S., Margulis, Katherine, Sullivan,
  Delaney~K., Toal, Georgia~G.  \emph{and others}}. (2017, April).
\newblock Oncogene {KRAS} activates fatty acid synthase, resulting in specific
  {ERK} and lipid signatures associated with lung adenocarcinoma.
\newblock {\em Proceedings of the National Academy of
  Sciences\/}~\textbf{114}(17), 4300--4305.

\bibitem[Guo \emph{and others}(2016)Guo, Wang, Feng, Bao, He, Bao, Hao and
  Wang]{guo_rapamycin_2016}
\textsc{Guo, Zhixin, Wang, Yanfeng, Feng, Xue, Bao, Chaogetu, He, Qiburi
  \emph{and others}}. (2016, January).
\newblock Rapamycin {Inhibits} {Expression} of {Elongation} of
  {Very}-long-chain {Fatty} {Acids} 1 and {Synthesis} of {Docosahexaenoic}
  {Acid} in {Bovine} {Mammary} {Epithelial} {Cells}.
\newblock {\em Asian-Australasian Journal of Animal
  Sciences\/}~\textbf{29}(11), 1646--1652.

\bibitem[Hoff(2019)Hoff]{hoff_smaller_2019}
\textsc{Hoff, Peter~D.} (2019, July).
\newblock Smaller \$p\$-values via indirect information.
\newblock {\em arXiv:1907.12589 [stat]\/}.
\newblock arXiv: 1907.12589.

\bibitem[Horn and Johnson(2008)Horn and Johnson]{horn_topics_2008}
\textsc{Horn, Roger~A. and Johnson, Charles~R.} (2008).
\newblock {\em Topics in matrix analysis\/}, 10. printing edition. Cambridge:
  Cambridge Univ. Press.
\newblock OCLC: 846124886.

\bibitem[Iorio \emph{and others}(2016)Iorio, Knijnenburg, Vis, Bignell, Menden,
  Schubert, Aben, Gonçalves, Barthorpe, Lightfoot, Cokelaer, Greninger, van
  Dyk, Chang, de~Silva, Heyn, Deng, Egan, Liu, Mironenko, Mitropoulos,
  Richardson, Wang, Zhang, Moran, Sayols, Soleimani, Tamborero, Lopez-Bigas,
  Ross-Macdonald, Esteller, Gray, Haber, Stratton, Benes, Wessels,
  Saez-Rodriguez, McDermott and Garnett]{iorio_landscape_2016}
\textsc{Iorio, Francesco, Knijnenburg, Theo~A., Vis, Daniel~J., Bignell,
  Graham~R., Menden, Michael~P.  \emph{and others}}. (2016, July).
\newblock A {Landscape} of {Pharmacogenomic} {Interactions} in {Cancer}.
\newblock {\em Cell\/}~\textbf{166}(3), 740--754.

\bibitem[Khan and Kaski(2014)Khan and Kaski]{calders_bayesian_2014}
\textsc{Khan, Suleiman~A and Kaski, Samuel}. (2014).
\newblock Bayesian {Multi}-view {Tensor} {Factorization}.
\newblock In: Calders, Toon, Esposito, Floriana, Hüllermeier, Eyke and Meo,
  Rosa (editors), {\em Machine {Learning} and {Knowledge} {Discovery} in
  {Databases}\/}, Volume 8724. Berlin, Heidelberg: Springer Berlin Heidelberg,
  pp.\  656--671.

\bibitem[Kory \emph{and others}(2018)Kory, Wyant, Prakash, uit~de Bos,
  Bottanelli, Pacold, Chan, Lewis, Wang, Keys, Guo and
  Sabatini]{kory_sfxn1_2018}
\textsc{Kory, Nora, Wyant, Gregory~A., Prakash, Gyan, uit~de Bos, Jelmi,
  Bottanelli, Francesca  \emph{and others}}. (2018, November).
\newblock {SFXN1} is a mitochondrial serine transporter required for one-carbon
  metabolism.
\newblock {\em Science\/}~\textbf{362}(6416), eaat9528.

\bibitem[Liao \emph{and others}(2017)Liao, Davoli, Leng, Li, Xu and
  Elledge]{liao_genetic_2017}
\textsc{Liao, Sida, Davoli, Teresa, Leng, Yumei, Li, Mamie~Z., Xu, Qikai and
  Elledge, Stephen~J.} (2017, January).
\newblock A genetic interaction analysis identifies cancer drivers that modify
  {EGFR} dependency.
\newblock {\em Genes \& Development\/}~\textbf{31}(2), 184--196.

\bibitem[Lim \emph{and others}(2009)Lim, Vaillant, Wu, Forrest, Pal, Hart,
  Asselin-Labat, Gyorki, Ward, Partanen, Feleppa, Huschtscha, Thorne, Fox, Yan,
  French, Brown, Smyth, Visvader and Lindeman]{lim_aberrant_2009}
\textsc{Lim, Elgene, Vaillant, François, Wu, Di, Forrest, Natasha~C, Pal,
  Bhupinder  \emph{and others}}. (2009, August).
\newblock Aberrant luminal progenitors as the candidate target population for
  basal tumor development in {BRCA1} mutation carriers.
\newblock {\em Nature Medicine\/}~\textbf{15}(8), 907--913.

\bibitem[McFarland \emph{and others}(2018)McFarland, Ho, Kugener, Dempster,
  Montgomery, Bryan, Krill-Burger, Green, Vazquez, Boehm, Golub, Hahn, Root and
  Tsherniak]{mcfarland_improved_2018}
\textsc{McFarland, James~M., Ho, Zandra~V., Kugener, Guillaume, Dempster,
  Joshua~M., Montgomery, Phillip~G.  \emph{and others}}. (2018, December).
\newblock Improved estimation of cancer dependencies from large-scale {RNAi}
  screens using model-based normalization and data integration.
\newblock {\em Nature Communications\/}~\textbf{9}(1), 4610.

\bibitem[Meyers \emph{and others}(2017)Meyers, Bryan, McFarland, Weir,
  Sizemore, Xu, Dharia, Montgomery, Cowley, Pantel, Goodale, Lee, Ali, Jiang,
  Lubonja, Harrington, Strickland, Wu, Hawes, Zhivich, Wyatt, Kalani, Chang,
  Okamoto, Stegmaier, Golub, Boehm, Vazquez, Root, Hahn and
  Tsherniak]{meyers_computational_2017}
\textsc{Meyers, Robin~M, Bryan, Jordan~G, McFarland, James~M, Weir, Barbara~A,
  Sizemore, Ann~E  \emph{and others}}. (2017, December).
\newblock Computational correction of copy number effect improves specificity
  of {CRISPR}–{Cas9} essentiality screens in cancer cells.
\newblock {\em Nature Genetics\/}~\textbf{49}(12), 1779--1784.

\bibitem[Pusapati \emph{and others}(2018)Pusapati, Kong, Patel, Krishnan,
  Sagner, Kinnebrew, Briscoe, Aravind and Rohatgi]{pusapati_crispr_2018}
\textsc{Pusapati, Ganesh~V., Kong, Jennifer~H., Patel, Bhaven~B., Krishnan,
  Arunkumar, Sagner, Andreas  \emph{and others}}. (2018, January).
\newblock {CRISPR} {Screens} {Uncover} {Genes} that {Regulate} {Target} {Cell}
  {Sensitivity} to the {Morphogen} {Sonic} {Hedgehog}.
\newblock {\em Developmental Cell\/}~\textbf{44}(1), 113--129.e8.

\bibitem[Ritchie \emph{and others}(2015)Ritchie, Phipson, Wu, Hu, Law, Shi and
  Smyth]{ritchie_limma_2015}
\textsc{Ritchie, Matthew~E., Phipson, Belinda, Wu, Di, Hu, Yifang, Law,
  Charity~W., Shi, Wei and Smyth, Gordon~K.} (2015, April).
\newblock limma powers differential expression analyses for {RNA}-sequencing
  and microarray studies.
\newblock {\em Nucleic Acids Research\/}~\textbf{43}(7), e47--e47.

\bibitem[Ritchie \emph{and others}(2007)Ritchie, Silver, Oshlack, Holmes,
  Diyagama, Holloway and Smyth]{ritchie_comparison_2007}
\textsc{Ritchie, M.~E., Silver, J., Oshlack, A., Holmes, M., Diyagama, D.,
  Holloway, A. and Smyth, G.~K.} (2007, October).
\newblock A comparison of background correction methods for two-colour
  microarrays.
\newblock {\em Bioinformatics\/}~\textbf{23}(20), 2700--2707.

\bibitem[Salloum \emph{and others}(2014)Salloum, Mukhopadhyay, Tung,
  Polonetskaya and Foster]{salloum_mutant_2014}
\textsc{Salloum, D., Mukhopadhyay, S., Tung, K., Polonetskaya, A. and Foster,
  D.~A.} (2014, March).
\newblock Mutant {Ras} {Elevates} {Dependence} on {Serum} {Lipids} and
  {Creates} a {Synthetic} {Lethality} for {Rapamycin}.
\newblock {\em Molecular Cancer Therapeutics\/}~\textbf{13}(3), 733--741.

\bibitem[Smyth and Speed(2003)Smyth and Speed]{smyth_normalization_2003}
\textsc{Smyth, Gordon~K and Speed, Terry}. (2003, December).
\newblock Normalization of {cDNA} microarray data.
\newblock {\em Methods\/}~\textbf{31}(4), 265--273.

\bibitem[Sondka \emph{and others}(2018)Sondka, Bamford, Cole, Ward, Dunham and
  Forbes]{sondka_cosmic_2018}
\textsc{Sondka, Zbyslaw, Bamford, Sally, Cole, Charlotte~G., Ward, Sari~A.,
  Dunham, Ian and Forbes, Simon~A.} (2018, November).
\newblock The {COSMIC} {Cancer} {Gene} {Census}: describing genetic dysfunction
  across all human cancers.
\newblock {\em Nature Reviews Cancer\/}~\textbf{18}(11), 696--705.

\bibitem[Sutskever \emph{and others}(2009)Sutskever, Tenenbaum and
  Salakhutdinov]{sutskever_modelling_2009}
\textsc{Sutskever, Ilya, Tenenbaum, Joshua~B. and Salakhutdinov, Ruslan~R}.
  (2009).
\newblock Modelling {Relational} {Data} using {Bayesian} {Clustered} {Tensor}
  {Factorization}.
\newblock In: Bengio, Y., Schuurmans, D., Lafferty, J.~D., Williams, C. K.~I.
  and Culotta, A. (editors), {\em Advances in {Neural} {Information}
  {Processing} {Systems} 22\/}. Curran Associates, Inc., pp.\  1821--1828.

\bibitem[Tsherniak \emph{and others}(2017)Tsherniak, Vazquez, Montgomery, Weir,
  Kryukov, Cowley, Gill, Harrington, Pantel, Krill-Burger, Meyers, Ali,
  Goodale, Lee, Jiang, Hsiao, Gerath, Howell, Merkel, Ghandi, Garraway, Root,
  Golub, Boehm and Hahn]{tsherniak_defining_2017}
\textsc{Tsherniak, Aviad, Vazquez, Francisca, Montgomery, Phil~G., Weir,
  Barbara~A., Kryukov, Gregory  \emph{and others}}. (2017, July).
\newblock Defining a {Cancer} {Dependency} {Map}.
\newblock {\em Cell\/}~\textbf{170}(3), 564--576.e16.

\bibitem[Wang \emph{and others}(2017)Wang, Yu, Hughes, Liu, Kendirli, Klein,
  Chen, Lander and Sabatini]{wang_gene_2017}
\textsc{Wang, Tim, Yu, Haiyan, Hughes, Nicholas~W., Liu, Bingxu, Kendirli, Arek
   \emph{and others}}. (2017, February).
\newblock Gene {Essentiality} {Profiling} {Reveals} {Gene} {Networks} and
  {Synthetic} {Lethal} {Interactions} with {Oncogenic} {Ras}.
\newblock {\em Cell\/}~\textbf{168}(5), 890--903.e15.

\bibitem[Ye(2005)Ye]{ye_generalized_2005}
\textsc{Ye, Jieping}. (2005, November).
\newblock Generalized {Low} {Rank} {Approximations} of {Matrices}.
\newblock {\em Machine Learning\/}~\textbf{61}(1-3), 167--191.

\end{thebibliography}

\appendix

\section{Appendix}

\subsection{Empirical Bayes estimation of variance parameters}

\subsubsection{Data partitioning}

The empirical Bayes strategy for estimation of the relevance of the historical data is as follows: first, estimate $\tilde{\sigma}^2$. Then obtain estimators of $\tau^2, \psi^2$ by maximizing the marginal likelihood of the data under the assumption that $\sigma_j^2 = \tilde{\sigma}^2$ for $j = 1, \dots, M$. The assumption of homoscedasticity eases the computations involved with evaluating the marginal likelihood of the data, as explained below. However, it should be noted that this assumption applies only to the sampling model assumed when calculating the value for $b_j^\text{FAB}$, not the (possibly) heteroscedastic sampling model that underlies the validity of the FAB and classical hypothesis tests. The FAB test is valid as long as $b_j^\text{FAB}$ is statistically independent of the test statistic $T_j$.

The independence of $b_j^\text{FAB}$ and $T_j$ can be accomplished by partitioning the available data into two non-overlapping index sets. Let $\mathcal{J}$ be an index set containing a randomly selected half of the entries in $\{1, \dots, M\}$. Then let
\begin{equation}
    \begin{aligned}
        \tilde{\sigma}^2 &=  \sum_{j \in \mathcal{J}} \left(S_{j}^2 / \tilde{\nu} \right) \left(n_{j}-1\right)
    \end{aligned}
\end{equation}
where $\tilde{\nu} = \sum_{j \in \mathcal{J}}(n_{j}-1)$. Given $\tilde{\sigma}^2$ corresponding to the index set $\mathcal{J}$, estimators $\tilde{\tau}^2$ and $\tilde{\psi}^2$ can be found by maximizing the marginal likelihood of the data $\bar{Y}_j$, $j \in \mathcal{J}$ induced by the combined sampling model, linking model, and hierarchical prior on the coefficients $\boldsymbol{\beta}$ (see the following section). For $\bar{Y}_j$, $j \notin \mathcal{J}$, estimators $\hat{\sigma}_j^2$ can be computed from entries $S_j$ where $j \notin \mathcal{J}$ such that under the null hypothesis $H_j : \theta_{j} = 0$, the statistic $T_j = \bar{Y}_{j} / \sqrt{\hat{\sigma}_j^2/n_{j}}$ is distributed as a $t_{\hat{\nu}}$ random variable. Hence, any $b_j^\text{FAB}$, computed from $\tilde{\sigma}^2$, $\tilde{\tau}^2$, and $\tilde{\psi}^2$ using data $\bar{Y}_j, S_j$, $j \in \mathcal{J}$ is independent of any $T_j$ where $j \notin \mathcal{J}$, and vice versa.

\subsubsection{Maximizing the marginal data likelihood}

Consider the sampling and linking model with the normal hierarchical prior distribution on $\boldsymbol{\beta}$ as in the main text
\begin{equation}
\begin{aligned}
    \bar{\mathbf{Y}} | \boldsymbol{\theta} &\sim N_M(\boldsymbol{\theta}, \text{diag}(\sigma_1^2 / n_1, \dots, \sigma_M^2 / n_M)) \\
    \boldsymbol{\theta} | \boldsymbol{\beta} &\sim N_M(\mathbf{X} \boldsymbol{\beta}, \tau^2 \mathbf{I}_M) \\
    \boldsymbol{\beta} &\sim N_q(0, \psi^2 \mathbf{I}_q)
\end{aligned}
\end{equation}
Under this model, the induced marginal density on $\bar{\mathbf{Y}}$ is
\begin{equation}
p(\bar{\mathbf{Y}}) = \frac{1}{\sqrt{2 \pi |\mathbf{A}|}} \exp \left(-\frac{\bar{\mathbf{Y}}^\top \mathbf{A}^{-1} \bar{\mathbf{Y}}}{2}\right)
\end{equation}
where the marginal covariance of $\bar{\mathbf{Y}}$ is 
\begin{equation}
\mathbf{A} = \psi^2 \mathbf{X} \mathbf{X}^\top + \tau^2 \mathbf{I}_M + \text{diag}(\sigma_1^2 / n_1, \dots, \sigma_M^2 / n_M)
\end{equation}
As long as the rows of $\mathbf{X}$ are not mutually orthogonal---which would make $\mathbf{X}\mathbf{X}^\top$ equal to the identity---variation due to $\psi^2$ and variation due to $\tau^2$ can be disambiguated. In fact, given values for the $\sigma_j^2$, maximum likelihood estimators for $\psi^2$ and $\tau^2$ are available, even when $p > n$. 

The log-likelihood of $\psi^2$ and $\tau^2$ under the marginal density for $\bar{\mathbf{Y}}$ is
\begin{equation}
\ell(\tau^2, \psi^2) = -\frac{1}{2} \left( \log 2\pi + \log |\mathbf{A}| + \bar{\mathbf{Y}}^\top \mathbf{A}^{-1} \bar{\mathbf{Y}} \right)
\end{equation}
The maximizer of $\ell(\tau^2, \psi^2)$ is not available in closed form. In addition, numerical optimization applied to the expression above requires computing the $M \times M$ matrix $\mathbf{A}$, which can become very large when the number of hypotheses is large. Making the approximation $\tilde{\sigma}^2 / n_j \approx \tilde{\sigma}^2 / \bar{n}$ for $j = 1, \dots, M$ eases much of this computational burden. Letting $\mathbf{X}\mathbf{X}^\top = \mathbf{Q} \boldsymbol{\Lambda} \mathbf{Q}^\top$ be the orthogonal eigendecomposition of the symmetric matrix $\mathbf{X}\mathbf{X}^\top$, $\mathbf{A}$ can now be written as
\begin{equation}
\mathbf{A} = \mathbf{Q} (\tau^2 \mathbf{I}_M + \psi^2 \boldsymbol{\Lambda} + (\tilde{\sigma}^2 / \bar{n}) \mathbf{I}_M)\mathbf{Q}^\top
\end{equation}
from which it is clear that the eigenvalues of $\mathbf{A}$ are 
\begin{equation}
\tau^2 + \psi^2 \lambda_j + \tilde{\sigma}^2 / \bar{n}, ~~~~~ j = 1, \dots, M
\end{equation}
where $\lambda_i$ is the $i^\text{th}$ eigenvalue of $\mathbf{X}$. The first $q$ eigenvectors of $\mathbf{X}\mathbf{X}^\top$ can be obtained by taking the singular value decomposition of $\mathbf{X}$. If $q \geq M$, then these are all that is needed to evaluate the log-likelihood. If $q < M$, then the $M - q$ remaining eigenvectors do not need to be computed explicitly. Since the columns of $\mathbf{Q}$ form an orthonormal eigenbasis for $\mathbb{R}^M$, these eigenvectors correspond to the directions that are orthogonal to the first $q$ eigenvectors of $\mathbf{X}\mathbf{X}^\top$. In addition, the associated eigenvalues of these directions are $0$. Hence, the component of the quadratic form $\bar{\mathbf{Y}} \mathbf{A}^{-1} \bar{\mathbf{Y}}$ corresponding to these directions has total norm 
\begin{equation}
\frac{||\bar{\mathbf{Y}}||_2^2 - \sum_{j=1}^q ||\mathbf{Q}_j^\top \bar{\mathbf{Y}}||_2^2}{\tau^2 + \tilde{\sigma}^2 / \bar{n}}.
\end{equation}
The maximum likelihood estimators for $\tau^2, \psi^2$ are thus found by numerically finding $\tau^2, \psi^2$ that maximize
\begin{equation}
     -\sum_{j=1}^q \left[ \log (\tau^2 + \psi^2 \lambda_j + \tilde{\sigma}^2 / \bar{n}) + \frac{||\mathbf{Q}_j^\top \bar{\mathbf{Y}}||_2^2}{\tau^2 + \psi^2 \lambda_j + \tilde{\sigma}^2 / \bar{n}} \right] - \mathbb{I}[q < M] \frac{||\bar{\mathbf{Y}}||_2^2 - \sum_{j=1}^q ||\mathbf{Q}_j^\top \bar{\mathbf{Y}}||_2^2}{\tau^2 + \tilde{\sigma}^2 / \bar{n}}
\end{equation}
Finally, note that the Kronecker structure of $\mathbf{X}$ specific to the tensor model described in the main text can be exploited to calculate the eigenvalues and eigenvectors of $\mathbf{X}\mathbf{X}^\top$ by computing the singular value decompositions of $\mathbf{U}$ and $\mathbf{V}$.

\subsection{Result for leave-one-out estimation}

The following result uses linear algebra identities to obtain expressions for the leave-one-out (LOO) estimators
\begin{equation}\label{eq:prior_params}
    \tilde{m}_j = \mathbf{X}_j^\top \mathbf{G}_{-j}^{-1} \mathbf{X}_{-j}^\top \bar{\mathbf{Y}}_{-j}~~~~~\tilde{v}_j = \left(\frac{W_{jj}}{W_{jj} + \tau^2}\right)^2\mathbf{X}_j^\top \mathbf{G}_{-j}^{-1}\mathbf{X}_j + \tau^2.
\end{equation}
The LOO estimators can be written in terms of the corresponding estimators based on the full model. The LOO estimators corresponding to the rows of a matrix can therefore be calculated with only one matrix inversion operation on the $q \times q$ matrix used to fit the full model. Recall from the main text that $\mathbf{G} = \mathbf{X}^\top \mathbf{H} \mathbf{X} + (1/\psi^2)\mathbf{I}_q$ and note that
\begin{equation}
    \mathbf{X}^\top \mathbf{H} \mathbf{X} + (1/\psi^2)\mathbf{I}_q = \sum_{j=1}^M H_{jj} \mathbf{X}_j \mathbf{X}_j^\top + (1/\psi^2)\mathbf{I}_q
\end{equation}
so
\begin{equation}
    \begin{aligned}
        \mathbf{G}_{-j}^{-1} &= \left(\mathbf{X}_{-j}^\top \mathbf{H}_{-j} \mathbf{X}_{-j} + (1/\psi^2)\mathbf{I}_q\right)^{-1} \\
        &= \left(\mathbf{X}^\top \mathbf{H} \mathbf{X} + (1/\psi^2)\mathbf{I}_q -  H_{jj} \mathbf{X}_j \mathbf{X}_j^\top \right)^{-1}. \\
    \end{aligned}
\end{equation}
From here, use the Woodbury matrix inversion identity to obtain
\begin{equation}\label{eq:gsub}
    \mathbf{G}_{-j}^{-1} = \mathbf{G}^{-1} + \left( \frac{H_{jj}}{1 - H_{jj} \mathbf{X}_j^\top \mathbf{G}^{-1} \mathbf{X}_j} \right) \mathbf{G}^{-1} \mathbf{X}_j \mathbf{X}_j^\top \mathbf{G}^{-1}.
\end{equation}
Hence,
\begin{equation}\label{eq:eff_m}
    \begin{aligned}
        \tilde{m}_j &= \mathbf{X}_j^\top \mathbf{G}_{-j}^{-1} \mathbf{X}_{-j}^\top \bar{\mathbf{Y}}_{-j} \\
        &= \mathbf{X}_j^\top ( \mathbf{G}_{-j}^{-1} \mathbf{X}^\top \mathbf{H} \mathbf{\bar{Y}} - \mathbf{G}_{-j}^{-1} \mathbf{X}_j H_{jj} \bar{Y}_j ) \\
    \end{aligned}
\end{equation}
Substituting \ref{eq:gsub} into \ref{eq:eff_m} and simplifying yields
\begin{equation}
    \begin{aligned}
        \tilde{m}_j = \tilde{\theta}_j - \frac{H_{jj} \mathbf{X}_j^\top \mathbf{G}^{-1} \mathbf{X}_j }{1 - H_{jj} \mathbf{X}_j^\top \mathbf{G}^{-1} \mathbf{X}_j} (\bar{Y}_j - \tilde{\theta}_j)
    \end{aligned}
\end{equation}
where $\tilde{\theta}_j = \mathbf{X}_j^\top \mathbf{G}^{-1} \mathbf{X}^\top \bar{\mathbf{Y}}$. Performing a similar substitution into the expression for $\tilde{v}_j$ yields
\begin{equation}
    \tilde{v}_j = \left(\frac{W_{jj}}{W_{jj} + \tau^2}\right)^2 \frac{H_{jj} \mathbf{X}_j^\top \mathbf{G}^{-1} \mathbf{X}_j }{1 - H_{jj} \mathbf{X}_j^\top \mathbf{G}^{-1} \mathbf{X}_j} + \tau^2
\end{equation}

\subsection{Inference and Gibbs sampling in the tensor probability model}

A priori, it is assumed that the cell line representations $\mathbf{U}_l$ are independently distributed as $N_{d_U}(\mathbf{0}, \mathbf{I})$ and the gene representations $\mathbf{V}_g$ are independently distributed as $N_{d_V}(\mathbf{0}, \mathbf{I})$. The matrices $\mathbf{B}_k$ are vectorized and the $\text{vec}(\mathbf{B}_k)$ are given independent flat priors $N_{d_U \times d_V}(0, \infty \mathbf{I})$. Finally, it is assumed that each precision variable $1/\tau_k^2$ is drawn independently from $G(1 / 2, 1 / 2)$ and each $\mu_k$ is drawn independently from $N(0, 1)$. 

A Gibbs sampling algorithm is used to obtain samples from the posterior distribution of the parameters in the tensor probability model. In each Gibbs step, the cell line features are sampled using the full conditional distribution
\begin{equation}
	\begin{aligned}
		&\mathbf{U}_l | \boldsymbol{\Theta}, \mathbf{V}, \mathbf{B}, \boldsymbol{\tau}^2 \sim N_{d_U} \left( \eta_l^* , \left[ \Lambda_l^* \right]^{-1} \right)
	\end{aligned}
\end{equation}
with
\begin{equation}
    \begin{aligned}
    	&\Lambda_l^* = \mathbf{I} + \sum_{g,k} (1 / \tau_k^2) \mathbf{B}_k \mathbf{V}_g {\mathbf{V}_g}^\top {\mathbf{B}_k}^\top \\
		&\eta_i^* = \left[ \Lambda_l^* \right]^{-1} \left(\sum_{g,k}  (1/\tau_k^2)(\theta_{lgk} - \mu_k) \mathbf{B}_k \mathbf{V}_g \right).
    \end{aligned}
\end{equation}
Analogous expressions are used to update the gene features $\mathbf{V}_g$. The each $\text{vec}(\mathbf{B}_k)$ is sampled from the conditional distribution
\begin{equation}
	\begin{aligned}
		&\text{vec} (\mathbf{B}_k) | \boldsymbol{\Theta}, \mathbf{U}, \mathbf{V}, \boldsymbol{\tau}^2 \sim N_{d_V \times d_U} \left( \boldsymbol{\xi}_k^* , \left[ \boldsymbol{\Psi}_k^* \right]^{-1} \right) \\
		&\boldsymbol{\Psi}_k^* = (1 / \tau_k^2) \sum_{i, j} \left[ \left(\mathbf{V}_g {\mathbf{V}_g}^\top \right) \otimes \left(\mathbf{U}_l {\mathbf{U}_l}^\top \right) \right] \\
		&\boldsymbol{\xi}_k^* = \left[ \boldsymbol{\Psi}_k^* \right]^{-1} \left( (1 / \tau_k^2) \sum_{l, g} (\theta_{lgk} - \mu_k) (\mathbf{V}_g \otimes \mathbf{U}_l) \right).
	\end{aligned}
\end{equation}
Finally, $\tau_k^2$ is sampled according to
\begin{equation}
	\begin{aligned}
		&1/ \tau_k^2 | \boldsymbol{\Theta}, \mathbf{U}, \mathbf{V}, \mathbf{B} \sim G \left( a_k^*, b_k^* \right) \\
		&a_k^* = \frac{N M + 1}{2} \\
		&b_k^* = \frac{\sum_{l, g} \left[ \theta_{lgk} - \mu_k - {\mathbf{U}_l}^\top \mathbf{B}_k \mathbf{V}_g \right]^2 + 1}{2}
	\end{aligned}
\end{equation}
and $\mu_k$ according to
\begin{equation}
	\begin{aligned}
		&\mu_k | \boldsymbol{\Theta}, \mathbf{U}, \mathbf{V}, \mathbf{B}, \tau_k^2 \sim N \left( \gamma_k^*, \left[\phi_k^*\right]^{-1} \right) \\
		&\gamma_k^* = \left[\phi_k^*\right]^{-1}\left(\sum_{i, j}(1 / \tau_k^2) (\theta_{lgk} - \mathbf{U}_l^\top \mathbf{B}_k \mathbf{V}_g)\right) \\
		&\phi_k^* = (NM / \tau_k^2) + 1
	\end{aligned}
\end{equation}

Inference for the biological effects $\theta_{lgk}$ proceeds as follows: for a modality $k$ with data modeled with the normal sampling model, $\theta_{lgk} = Y_{lgk}$ at each Gibbs iteration. For a modality $k$ with binary data, the $\theta_{lgk}$ have full conditional distributions given by
\begin{equation}
\begin{aligned}
	\theta_{lg k'} | \mathbf{Y}, \mathbf{U}, \mathbf{V}, \mathbf{B} &\sim \left\{
        \begin{array}{ll}
            N_+ (\mu_{k'} + \mathbf{U}_l^\top \mathbf{B}_{k'} \mathbf{V}_g, 1) & \quad \text{if}~~Y_{lgk'} = 1 \\
            \\
            N_- (\mu_{k'} + \mathbf{U}_l^\top \mathbf{B}_{k'} \mathbf{V}_g, 1) & \quad \text{if}~~Y_{lgk'} = 0 \\
        \end{array}
    \right.
    \end{aligned}
\end{equation}
where $N_\pm$ denotes the normal distribution truncated from below and above, respectively, at $0$. For a modality $k$ with strictly positive continuous data, the $\theta_{lgk}$ have full conditional distributions
\begin{equation}
\begin{aligned}
	\theta_{lg k} | \mathbf{Y}, \mathbf{U}, \mathbf{V}, \mathbf{B}, \tau_{k}^2 &\sim \left\{
        \begin{array}{ll}
            Y_{lgk} & \quad \text{if}~~Y_{lgk} > 0 \\
            \\
            N_- (\mu_{k} + \mathbf{U}_l^\top \mathbf{B}_{k} \mathbf{V}_g, \tau_{k}^2) & \quad \text{if}~~Y_{lgk} = 0 \\
        \end{array}
    \right.
    \end{aligned}
\end{equation}

Missingness in large genomics corpora often has to do with availability of cell line samples or sequencing library size. Therefore, it is reasonable to assume that missing data are missing at random (MAR). Under a MAR assumption, marginalizing out missing data is accomplished by treating the missing data like model parameters within the Gibbs sampling steps: at each iteration, a new $Y_{lgk}$ is drawn according to its sampling model given the other model parameters. For the normal model, this means $Y_{lgk}$ is drawn from
\begin{equation}
    Y_{lgk} | \mathbf{U}, \mathbf{V}, \mathbf{B}, \tau_{k}^2 \sim N (\mu_{k} + \mathbf{U}_l^\top \mathbf{B}_{k} \mathbf{V}_g, \tau_{k}^2).
\end{equation}
In the probit model, $Y_{lgk}$ is drawn from
\begin{equation}
    Y_{lgk} | \mathbf{U}, \mathbf{V}, \mathbf{B}, \tau_{k}^2 \sim B (\Phi(\mu_{k} + \mathbf{U}_l^\top \mathbf{B}_{k} \mathbf{V}_g))
\end{equation}
and in the tobit model $Y_{lgk}$ is drawn from
\begin{equation}
    Y_{lgk} | \mathbf{U}, \mathbf{V}, \mathbf{B}, \tau_{k}^2 \sim  N_+ (\mu_{k} + \mathbf{U}_l^\top \mathbf{B}_{k} \mathbf{V}_g, \tau_{k}^2)
\end{equation}

\subsection{Identifiability and Procrustes alignment}

The parameters $\mathbf{U}, \mathbf{B}, \mathbf{V}$ in the tensor factorization model are not identifiable, since for any square orthogonal matrices $\mathbf{Q}, \mathbf{P}$ and diagonal scaling matrices $\mathbf{L}, \mathbf{J}$ the following holds
\begin{equation}
    \mathbf{U} \mathbf{B}_k \mathbf{V}^\top = \mathbf{U}\mathbf{LQ}^\top\mathbf{QL}^{-1}\mathbf{B}_k\mathbf{J}^{-1}\mathbf{P}^\top\mathbf{PJ}\mathbf{V}^\top
\end{equation}
In MCMC sampling algorithms, like Gibbs sampling, it is common to observe that such rotation and scale ambiguity leads to misalignment of samples from disparate parts of the Markov Chain. Hence, before taking posterior summaries, it is necessary to align the samples of the matrices $\mathbf{U}, \mathbf{B}$ and $\mathbf{V}$ to a common rotation and scale. This is accomplished by a Procrustes alignment procedure, followed by a scaling of each of the aligned $\mathbf{U}$ and $\mathbf{V}$ samples to have unit column norm. The scaling and rotation operations applied to each sample of $\mathbf{U}$ and $\mathbf{V}$ are then passed to the samples of $\mathbf{B}_k$, so that the overall matrix product is preserved---that is, if $\mathbf{U}^*$, $\mathbf{V}^*$, and $\mathbf{B}_k^*$ are the aligned counterparts to $\mathbf{U}$, $\mathbf{V}$ and $\mathbf{B}_k$, then
\begin{equation}
    \mathbf{U}^*\mathbf{B}_k^*\mathbf{V}^{*\top} = \mathbf{U} \mathbf{B}_k \mathbf{V}^\top
\end{equation}
for every sample.

\subsection{Choice of cancer cell lines and genes for inclusion in tensor probability model}

Cancer cell line historical feature profiles were inferred for all cancer cell lines that appeared in at least one of the four cancer genomics datasets from the Dependency Map collection. This totaled $1209$ cancer cell lines. The list of genes for which historical information profiles were derived was obtained by taking the union of the following sets of genes: (1) genes in the COSMIC cancer gene census \citep{sondka_cosmic_2018}, (2) genes in the \cite{kory_sfxn1_2018} study, (3) genes in the \cite{wang_gene_2017} focused screen, (4) the top $1,000$ genes as ranked by variance of gene expression profile, and (5) the top $1,000$ genes as ranked by sum of squared gene expression values. In total, this list contained $4570$ genes.

\end{document}